\documentclass[a4paper,fleqn,usenatbib,useAMS]{mnras}
\usepackage[dvips]{graphicx}
\usepackage{color}
\usepackage{times}
\usepackage{natbib}
\usepackage{setspace}
\usepackage{amsmath}
\usepackage{amsfonts}
\usepackage{amssymb}
\usepackage{multirow}
\usepackage{multicol}
\usepackage{aas}
\usepackage{booktabs}
\usepackage{adjustbox}
\usepackage{graphics}
\usepackage{float}
\usepackage{blindtext}
\usepackage{enumitem}
\usepackage{ragged2e}
\usepackage{etoolbox}
\makeatletter
\usepackage{hyperref}
\usepackage{bm}		
\usepackage[T1]{fontenc}
\usepackage{ae,aecompl}

\def\spb{\smallskip\par\noindent$\bullet\;$}

\newif\ifAMStwofonts
\AMStwofontstrue
\definecolor{red}{rgb}{1,0.,0.}

\title[Galaxy protocluster at $z=5.2$]{Probing the existence of a rich galaxy overdensity at $z=5.2$}

\author[Rosa Calvi et al.]{Rosa Calvi$^{1,2}$\thanks{E-mail: rosa.calvi@gmail.com},
 Helmut Dannerbauer$^{1,2}$, Pablo Arrabal Haro$^{1,2}$, Jos\'e M. Rodr\'iguez \newauthor Espinosa$^{1,2}$, Casiana Mu\~noz-Tu\~n\'on$^{1,2}$, Pablo G. P\'erez Gonz\'alez$^{3}$, Stefan Geier$^{1,4}$\\
$^1$ Instituto de Astrof\'isica de Canarias, E-38205 La Laguna, Spain \\ 
 $^2$ Depto. de Astrof\'isica, Universidad de La Laguna, E-38206 La Laguna, Spain\\
$^3$ Centro de Astrob\'iolog\'ia, Departamento de Astrof\'isica, CSIC-INTA, Ctra. de Ajalvir km 4, E-28850—Torrej\'on de Ardoz, Madrid, Spain\\
$^4$ Gran Telescopio Canarias (GRANTECAN), Cuesta de San José s/n, 38712 Breña Baja, La Palma, Spain
}

\pubyear{2020}

\begin{document}

\label{firstpage}
\pagerange{\pageref{firstpage}--\pageref{lastpage}}
\maketitle

\begin{abstract}
We report the results of a pilot spectroscopic program of a region at $z=5.2$ in the GOODS-N field containing an overdensity of galaxies around the well-known submillimeter galaxy HDF850.1. We have selected candidate cluster members from the optical 25 medium-band photometric catalog of the project SHARDS (Survey for High-z Absorption Red and Dead Sources). 17 rest-frame UV selected galaxies (LAEs and LBGs) with 5.15<$z_{phot}$<5.27, candidates to be physically associated with the overdensity, have been observed with the instrument OSIRIS at the GranTeCan telescope. 13 out of these 17 (76 per cent) sources have secure spectroscopic confirmations via the Ly$\alpha$ line at the redshift of the galaxy protocluster PCl$-$HDF850.1, demonstrating the high reliabilty of our photometric redshift method. 10 out of 13 sources are newly confirmed members. Thus, we increase the number of confirmed members in this overdensity from 13 to 23 objects. In order to fully characterize this structure we combined our dataset with the sample from the literature. Beside the SMG HDF850.1, none of the 23 spectroscopically confirmed members are bright in the far-infrared/submm wavelength regime (SFR$_{\rm IR}<$ few hundred M$_{\odot}$ yr$^{-1}$). The clustering analysis of the whole sample of 23 confirmed members reveals four distinct components in physical space in different evolutionary states, within $\Delta z<0.04$ from the central region hosting SMG HDF850.1. The halo mass of the whole structure at $z=5.2$, estimated by a variety of methods, range between $2-8\times 10^{12}M_{\odot}$. The comparison with literature suggests a large scale assembly comparable to the formation of a central Virgo-like cluster at z=0 with several satellite components which will possibly be incorporated in a single halo if the protocluster is the progenitor of a
more massive Coma-like cluster ($>10^{15}M_{\odot}$).

\end{abstract}


\section{Introduction}

Over the last decades, observations and theoretical predictions of galaxy clusters, the most massive gravitationally bound structures, provided very important clues on the formation and evolution of galaxies \citep[e.g.,][]{Kravtsov2012}. In particular, it is fairly well established that all physical processes coming into play after a galaxy has become part of a dense structure, like a group or a cluster, have a significant impact on the morphological transitions from an active (star-forming) phase to a more passive and chemical evolved phase \citep{Kauffmann04,Andreon04,Baldry06,Weinmann06,Balogh11}. However, quantifying the impact of environment on galaxy properties, especially those closely related to the galaxy-intrinsic conditions, e.g., mass of the halo, formation time \citep{Wang13,DeLucia07}, is still pending.

In the last years, overdense regions at high redshifts \citep[here after called `galaxy protoclusters';][]{Overzier16} that evolve into galaxy clusters observed in the local Universe such as Coma or Virgo has drawn more and more attention. The observations of these regions are important not only to investigate the early stage of galaxy formation and the subsequent evolution but also to provide information on the cosmology-dependent evolving density fluctuation peaks, setting important constraints on the cosmological parameters \citep{Kravtsov2012}. 

Many efforts have been done in recent years for finding these overdense structures at $z\sim1.5-2.5$, the epoch when the cosmic star formation rate (density), the accretion rate of gas feeding black holes and galaxy mergers reached their peak activity before being subsequently suppressed \citep[e.g.,][]{Dickinson03,Hopkins04,Madau2014,Muldrew15,Chiang17,Muldrew18}. At these redshifts, galaxy clusters and protoclusters were in process of being transformed from dynamical overdensities to the more relaxed systems we see today. Indeed they are excellent laboratories for studying galaxy assembly and particularly the impact of environment on the galaxy populations belonging to these systems. 

A variety of techniques have been established to search for galaxy overdensities such galaxy clusters and protoclusters (galaxy clusters in formation) as:  the red-sequence technique \citep[e.g.,][]{Gladders05,Goto08,Andreon09,Muzzin09,Wilson09}, the massive galaxies \citep[e.g.,][]{Daddi2009,Gobat2011}, the gaseous component in X-rays emission  \citep[e.g.,][]{Rosati02,Rosati04,Mullis05,Stanford06}, the strong absorption in the Ly$\alpha$ forest \citep[e.g.,][]{Cai2016,Miller19}, the weak-lensing shear selection \citep[e.g.,][]{Wittman06,Gavazzi07,Shan12,Jeffrey18}, the Sunyaev-Zeldovich signatures \citep[e.g.,][]{Staniszewski09, Mantz14}, emission lines such as Ly$\alpha$ and H$\alpha$ through narrow-band imaging and subsequent spectroscopy \citep[e.g.,][]{Kurk2000,Pentericci2000,Kurk2004a,Kurk2004b}, the Lyman break  features in the galaxy colour \citep[e.g.,][]{Miley04,Toshikawa18} and the
dust emission of star-forming galaxies \citep[DSFGs; e.g.,][]{Dannerbauer14,Planck,Greenslade19}. 

Successful methods usually search for overdensities around distant, massive galaxies used as signposts such as: QSOs \citep[e.g.,][]{Djorgovski03,Wold03,Stiavelli05,Kashikawa07,Overzier09,Stevens10,Falder11,Matsuda11,Trainor12,Husband13,Morselli14,Adams15,Hennawi15}, high-z radio galaxies  \citep[HzRGs; e.g.,][]{Pascarelle96,LeFevre96,Kurk2000,Pentericci2000,Venemans04,Venemans05,Venemans07,Kajisawa06,Kuiper11,Hatch11a,Hatch11b,Hayashi12,Cooke14,Dannerbauer14} or submillimeter galaxies \citep[SMGs; e.g.,][]{Riechers14,Casey15,Casey16,Pavesi18,Lacaille19}. 

Especially, the search for rest-frame signatures such as the Ly$\alpha$ line and/or Lyman-break around possible signposts of large-scale structures had been very successful to find protoclusters from the cosmic noon to cosmic dawn. Observational evidence for indications of protoclusters were found at $z=2.3$ and $z=3.09$ in the field of QSOHS1700+64 and SSA22 surveys \citep{Steidel98,Steidel00,Steidel05} using rest-frame UV spectroscopy of a sample of candidate high redshift galaxies selected on the basis of the Lyman break. At $z=2.9$ and $z=3.3$ the large spectroscopic survey VIMOS Ultra-Deep Survey (VUDS) found two exceptionally overdense region in the COSMOS field \citep{Cucciati14,Lemaux14}. Several structures have been also identified in the High-Redshift(Z) Emission Line Survey (HiZELS, \citet{Sobral13a} at three redshifts, $z=0.8$, 1.47 and 2.23 \citep{Cochrane18}. At higher redshift, \citet{Ouchi05} and \citet{Jiang18} spectroscopically confirmed luminous Ly$\alpha$ emission galaxy candidates tracing a massive protocluster at $z=5.7$ in the Subaru/XMM-Newton Deep Field (SXDS). More recently, there are confirmations of a protocluster in the SXDS at $z=6.5$ by using OSIRIS/GTC \citep{Chanchaiworawit19,Calvi19} which produces a remarkable number of ionising continuum photons capable of ionising a large bubble \citep{Espinosa20}. This epoch is close to the full reionisation of the Universe. Even though the faint primordial galaxies (being responsible for the reionization) can be difficult to observe, their location in groups or bubbles of ionised gas can enhance their visibility. Finally, \citet{Harikane19} using Keck/DEIMOS and Gemini/GMOS spectroscopy and \citet{Castellano2018} with VLT observations found overdensities at $z=6.6$ and $z=7$. 

When using SMGs as signposts of overdensities, the following open questions are adressed: i) how many of these dusty starbursts --- the progenitors of elliptical galaxies dominating local galaxy clusters -- are members of such a structure and ii) how much this source population contributes to the total star-formation densities of individual structures. In the past years, several structures containing SMGs have been reported \citep[e.g.,][]{Casey15,Capak11,Clements14,Clements16,Dannerbauer14,Harikane19,Hung16,Ivison13,Kato16,Lewis18,Miller18,Oteo18,Pavesi18}. 

In \citet[][from now on AH18]{Arrabal18} we presented a search for both Ly$\alpha$ emitters (LAEs) and Lyman Break Galaxies (LBG) in the GOODS North field (GOODS-N). In particular, we have found 55 sources \citep{Arrabal18} that may belong to an already known overdensity at $z\sim5.2$ \citep[][from now on W12; see section 2 of this manuscript for details]{Walter12} physically related to the well-know submillimeter galaxy HDF~850.1 \citep[e.g.,][]{Hughes98,Dunlop04,Walter12,Neri14} at $z=5.183$. This paper presents the results of the spectroscopic follow-up of 17 candidate protocluster members through multi-object spectroscopy (MOS) with OSIRIS (the Optical System for Imaging and low-Intermediate-Resolution Integrated Spectroscopy) at the GTC (Gran Telescopio Canarias). In Section 2 we discuss the sample selected from the SHARDS survey and in Section 3 we describe the OSIRIS MOS observations and data reduction. In Section 4 we present the results, followed by a detailed analysis in Section 5 and 6. In section 7 we discuss our findings in an evolutionary contexts. Finally, we summarize the presented work. We adopt the $\Lambda$CDM concordant Universe model ($\Omega_{\Lambda}=0.7$, $\Omega_{M}=0.3$ and $h=0.7$) \citep{Bahcall99}. Magnitudes are given in the AB system \citep{Oke83}.

\section{Parent Sample}
Our sample is drawn from a catalogue of 1558 rest-frame UV-selected high-z galaxies presented in \citet{Arrabal18}. They carried out a systematic search of Ly$\alpha$ emitters and/or Lyman break galaxies from $z\sim3.35$ to $z\sim6.8$, using 25 medium-band filters from 500 to 941~nm of the SHARDS \citep[Survey for High-z Absorption Red and Dead Sources][]{Perez13}. This survey was conducted with OSIRIS \citep{Cepa2010}  at the GTC, covering an area of $\sim 130$ arcmin$^{2}$ in GOODS-North. Within the robust sample of 1558 sources, 528 LAEs and 1030 LBGs, 55 sources have photometric redshifts around $z=5.2$. This spike in the redshift distribution of these rest-frame UV selected galaxies seems to be associated with a galaxy protocluster (PCl$-$HDF850.1) presented by \citet{Walter12}.

Through the analysis of the environment around the well-known SMG HDF850.1, \citet{Walter12} discovered a galaxy overdensity physically related to this dusty starburst located at $z_{CO}=5.183$. This SMG is undetected even in deep {\it HST} imaging in the rest-frame UV/optical and only seen in far-infrared/submillmeter wavelength regime where the spectroscopic redshifts comes from. In total, through rest-frame UV spectroscopy, \citet{Walter12} identified additional 12 members over a narrow redshift range between $5.185<z<5.213$. 11 of these 13 members are recovered by our deep Ly$\alpha$ imaging presented in \citet{Arrabal18}. Due to its faintness in the rest-frame UV/optical, no Ly$\alpha$ emission is expected to be seen in HDF850.1, confirmed by our data. The second source at position R.A.: $12:36:39.8$, Decl.: $+62:09:49.1$ and redshift $z=5.187$ is detected by SHARDS data but not selected as LAE and/or LBG.

In addition, 44 new potential members have been found. Unfortunately, beside one \citep[QSO at $z=5.18$][]{Barger02}, neither spectra nor physical information of the cluster members are published in \citet{Walter12}. 

\begin{figure*}
  
  \includegraphics[width=0.60\textwidth]{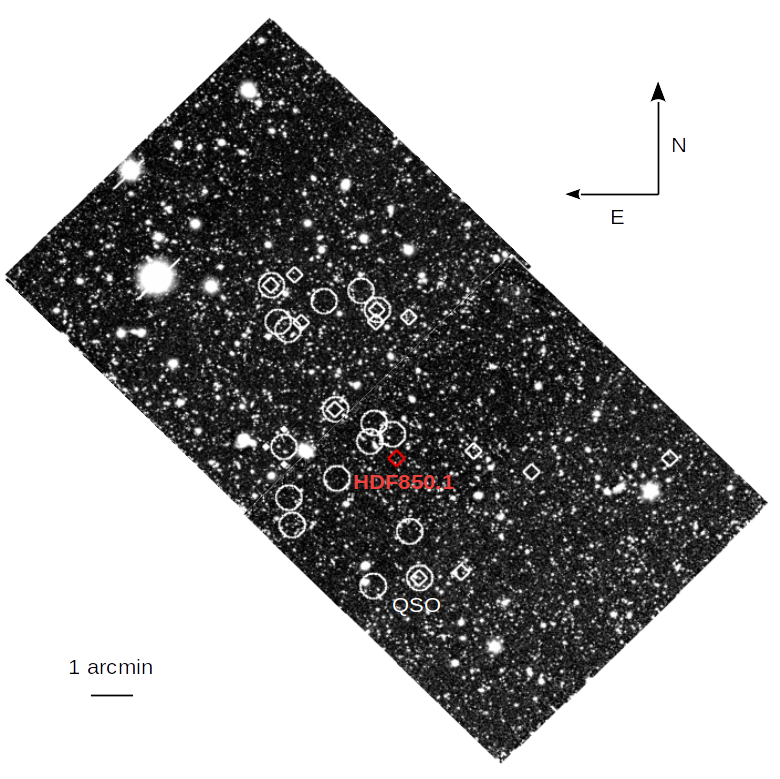}
  \caption{We show (candidate) members of the $z=5.2$ overdensity within the SHARDS area. The diamonds present the 13 members in \citet{Walter12} including the SMG~HD850.1 (red diamond). The circles present the 17 candidate members selected from AH18 for GTC OSIRIS MOS spectroscopy. Four sources have been already confirmed spectroscopically in \citet{Walter12}}.
  \label{field}
\end{figure*}

\begin{table*}
\caption{Observed targets}
\label{tab:detections}

	\begin{tabular}{@{}lcccccccl}
	\hline\hline
    Source & R.A. & Dec. & m$_{AB}$& $z_{\mathrm{phot}}$ & $z_{\mathrm{spec}}$ & Type & Grade & previous   \\
    &(J2000.0)&(J2000.0) & && &   & observations\\
   (1)&(2)&(3)&(4)&(5)&(6)&(7)&(8)&(9)\\
    \hline
    \hline
    \multicolumn{2}{l}{$z=5.2$ protocluster candidate members}\\
    SHARDS10005737 & 12:37:13.38 & 62:12:39.2  & 25.77$\pm$0.25 &5.26 & -- & LBG&C \\
    SHARDS10006357 & 12:37:15.63 &  62:16:23.6 & 25.75$\pm$0.18&5.19& 5.194& LAE&B&5.189 (W12)\\
    SHARDS10008210 & 12:37:14.50 & 62:15:32.4 & 26.11$\pm$0.26 &5.18 & 5.163 &LBG&B \\
    SHARDS10008850 & 12:36:55.39 & 62:15:48.8 & $>$26.92 &5.15 & -- & LBG&C& 5.190 (W12)\\
    SHARDS10010385 & 12:36:58.43 & 62:16:15.0 & 25.68$\pm$0.16 &5.16 & 5.195& LAE &A \\
    SHARDS10011501 & 12:37:05.52 & 62:16:01.3 & 25.50$\pm$0.08 &5.17 & 5.200& LAE&A \\
    SHARDS10018196 & 12:37:12.48 & 62:15:21.1 & 25.48$\pm$0.14 &5.17 & 5.195& LAE& A \\
    SHARDS20004537 & 12:36:47.96 & 62:09:41.4 & 22.74$\pm$0.01 &5.19 & 5.180 &LAE & A & 5.18, 5.186 (B02; W12) \\
    SHARDS20007254 & 12:37:12.80 & 62:11:32.0  & 26.07$\pm$0.34 &5.22& 5.218& LBG&B \\
    SHARDS20007459& 12:37:03.31 & 62:13:31.5 & 25.45$\pm$0,24 &5.27& 5.217 & LBG&A & 5.213 (W12)\\
    SHARDS20008702 & 12:37:12.08 & 62:10:54.1  & 25.11$\pm$0.11 &5.21 & 5.155 & LAE&A \\
    SHARDS20008777 & 12:36:56.70 & 62:09:30.5 & 25.03$\pm$0.14 &5.19 & 5.181 & LAE  &A \\
    SHARDS20008932 & 12:36:57.29 & 62:12:49.4 & 25.16$\pm$0.11 &5.23 & --& LAE& C \\
    SHARDS20010724 & 12:36:56.51 & 62:13:13.6 & 25.25$\pm$0.39 & 5.23  & 5.188&  LAE  &A \\
    SHARDS20011455 & 12:36:53.09 & 62:12:59.5 & 25.42$\pm$0.15 &5.23 & -- & LAE &C\\
    SHARDS20013107 & 12:36:49.79 & 62:10:45.0  & 25.22$\pm$0.12 &5.21 & 5.187 & LAE&A \\
    SHARDS20013448 & 12:37:03.61 & 62:11:58.5 & 24.48$\pm$0.11 &5.22 & 5.224 & LAE&A \\
    \hline
    \hline
    \end{tabular}
\raggedright \justify 
\textbf{Notes:}  
Column (1): SHARDS source name. Column (2): J2000.0 right ascension of the targets. Units of right ascension are hours, minutes, and seconds. Column (3): J2000.0 declination of targets. Units of declination are degrees, arcminutes, and arcseconds. Column (4): The magnitudes are given in the AB system \citep{Oke83}. The magnitudes are measured in the SHARDS filter F755w17.  Column (5): Photometric redshift measured from the photometry published in AH18, the typical error is  $\pm$0.07. Column (6): spectroscopic redshift, the typical error is accurate to the third digit. Column (7): Object type (LAE or LBG). Column (8): Quality of line detection, grade A is secure detection, B tentative detection and C no  no detection. Column (9): Known redshifts from literature.

\end{table*}

\section{Observations}
\subsection{GTC OSIRIS MOS spectroscopy}
The parent sample of possible overdensity members photometrically selected from \citet{Arrabal18} consists of 55 sources. We conducted a spectroscopic follow-up of a subset of candidates using the multi-object spectroscopy capability of OSIRIS at GTC in order to confirm their physically belonging to this overdensity. The primary goal was to maximise the number of sources in a single mask. The selection of the targets was made to achieve spatial uniformity. To optimise the success rate of this pilot program, we gave higher priority to the brightest sources. The mask was filled with so-called `bonus' galaxies (presumably not physically related to the $z=5.2$ overdensity) and was designed with the OSIRIS Mask Designer Tool (MD). The mask contained 24 objects, 17 LAE and LBG candidate cluster members around $z\sim 5.2$ and the seven `bonus' sources (which are beyond the scope of this work). Furthermore, four fiducial stars, without spectral superpositions from any pair of slits, are part of the mask in order to obtain the acquisition image. The OSIRIS field of view (FoV) is a mosaic of two CCDs with a small gap in-between. In principle, the FoV in the MOS mode is $7.5^{\prime}$ $\times$ $6.0^{\prime}$. We used the R2500I grism which produces a spectral resolution of $\sim$5~\AA{} at $\lambda=7450$\,\AA{}, with a 2$\times$2 binning, so the pixels scale is 0.254$^{\prime\prime}$/pixel. This setup allowed us to explore the spectral range between $\sim7330-10000$\,\AA{}. The observations (program ID: 167-GTC122/17B, PI: H. Dannerbauer) were obtained in service mode and taken at two different nights in February 2019. The number of observing blocks (OBs) were 4 and 5 during the first and second night, respectively. Each OB consisted of two frames of 1368 seconds each. The pure observing time was 6.9~hrs with a seeing of 0.7$^{\prime\prime}$ to 1.0$^{\prime\prime}$. The first night was dark and the second night was grey time. Only in the first night the conditions were photometric. The 2$^{\prime\prime}$ aperture circular holes were used for the fiducial stars and thus, for centring the mask with high precision. The science object slits had a slit length of 20$^{\prime\prime}$ and  a width of 1.2$^{\prime\prime}$. As the objects are very faint, we decided not to do dithering, thus avoiding to add extra noise to the data. The standard star used for flux calibration is Feige~66. In Figure~\ref{field} we show the locations of the 17 targets and the previously known 13 protocluster members from W12 around the SMG HDF850.1.

\subsection{Data Reduction}
The individual CCD images were reduced and combined using  IRAF routines \citep{iraf1986}. 
The data consisted of the scientific and calibration images. We subtracted the bias and normalised the flat for each OBs separately. We used a reference flat for each chip to remove detector and geometrical distortions. Then, the slits frames were aligned and then combined to obtain the master image to improve the S/N and remove the cosmic rays. Considering the the low signal-to-noise (S/N) and complex nature of the observed spectra, we decided to analyse separately the frames of the first and second night. Only after the extraction of the individual spectra, they were combined to maximise the signal-to-noise ratio (S/N). We separated the individual 24 slits, 9 in Chip1 and 15 in Chip2 and we trimmed off the edges in the science frames before aligning the spectra and performing the wavelength calibration. We performed the spectroscopic calibration using IRAF routines on the individual slit frames. Every slit is also contaminated by the emission spectrum of the Earth's atmosphere, thus we used the OH sky emission lines for the wavelength calibration of the spectra\footnote{We used the LRIS catalog of sky emission lines between 6400 and 10500 \AA{} from http://www.astrossp.unam.mx/~resast/standards/NightSky/skylines.html.}. Using the IRAF-NOAO package, we identified the corresponding wavelength sky emission lines within our slit frames. Fourth-order Chebyshev polynomials were fit along the spatial direction to provide a smooth master sky model to subtract it from the 2D spectrum. The sky subtraction process near emission lines can be problematic, therefore we also looked for significant peaks in the one-dimensional spectra which show an asymmetric line profile. We extracted a 1D spectrum by summing the flux along a window of 6 pixels ($\sim1.5''$). The size of the extraction window allowed us to optimize the flux counts of each emission line. Then, we found the centroid of the signal in the spectrum and we considered a window of 6 pixels around this centroid which contains the signal (+3 pixels on the right, -3 pixels on the left), we extracted the spectrum's lines in this window and, using IRAF tasks, we checked for the presence of an asymmetric profile. So, the total flux is the sum along the x-axis of this extracted spectrum. We focused on the 1D and 2D visual inspection in the wavelength range of $7300-7800\sim$\AA. Those spectra with a clear continuum break and/or emission line indicating a precise redshift were denoted as grade A, secure detection. On the opposite, if from visual inspection we were not able to obtain a clear detection, we performed a second iteration of the sky subtraction to look for a significant peak in the 2D spectrum. Sources detected in this way are grade B, tentative detection. A non-detection is category C. These are objects for which the resulting spectra have very low signal-to-noise ratio (S/N). During this process we found that not all science frames were useful. This was caused by non-photometric conditions of the second night, which was taken in grey and not in dark time. For this reason we discarded all the 10 science frames of the second night in which we could not even detect the brightest sources. In the end, we used a total of 8 science frames to obtain the highest S/N 2D spectra for our observed targets.

The spectra were flux calibrated using the spectroscopic standard star Feige~66 observed during the same observing run/night. The spectroscopic standard was used to determine the response curve of the spectrograph, and in turn was used to flux-calibrate the spectra of our targets. First, we reduced and calibrated in wavelength the spectroscopic standard star in the same manner as the science frames. However, since the standard star was taken with a wider slit width (2.5$^{\prime\prime}$), we used single slit flats for the standard from the OSIRIS webpage\footnote{http://www.gtc.iac.es/instruments/osiris/}. Then, we extracted the 1D spectrum of the standard star using the task IRAF/APALL. Knowing the exposure time and the extinction correction, we matched the 1D spectrum of the spectroscopic standard in units of ADU s$^{-1}$ to the available flux density catalogue of \citet{Oke90}, tabulated in the $onedstds$ directory of IRAF. Thus, we obtained the flux transformation from ADU s$^{-1}$ to erg s$^{-1}$ cm$^{-2}$ \AA{}$^{-1}$ to calibrate the science frames. The typical accuracy of the obtained spectroscopic redshifts is to the third digit.

\begin{figure*}

\centering
  \begin{tabular}{@{}cccc@{}}
   
    \includegraphics[width=.425\textwidth]{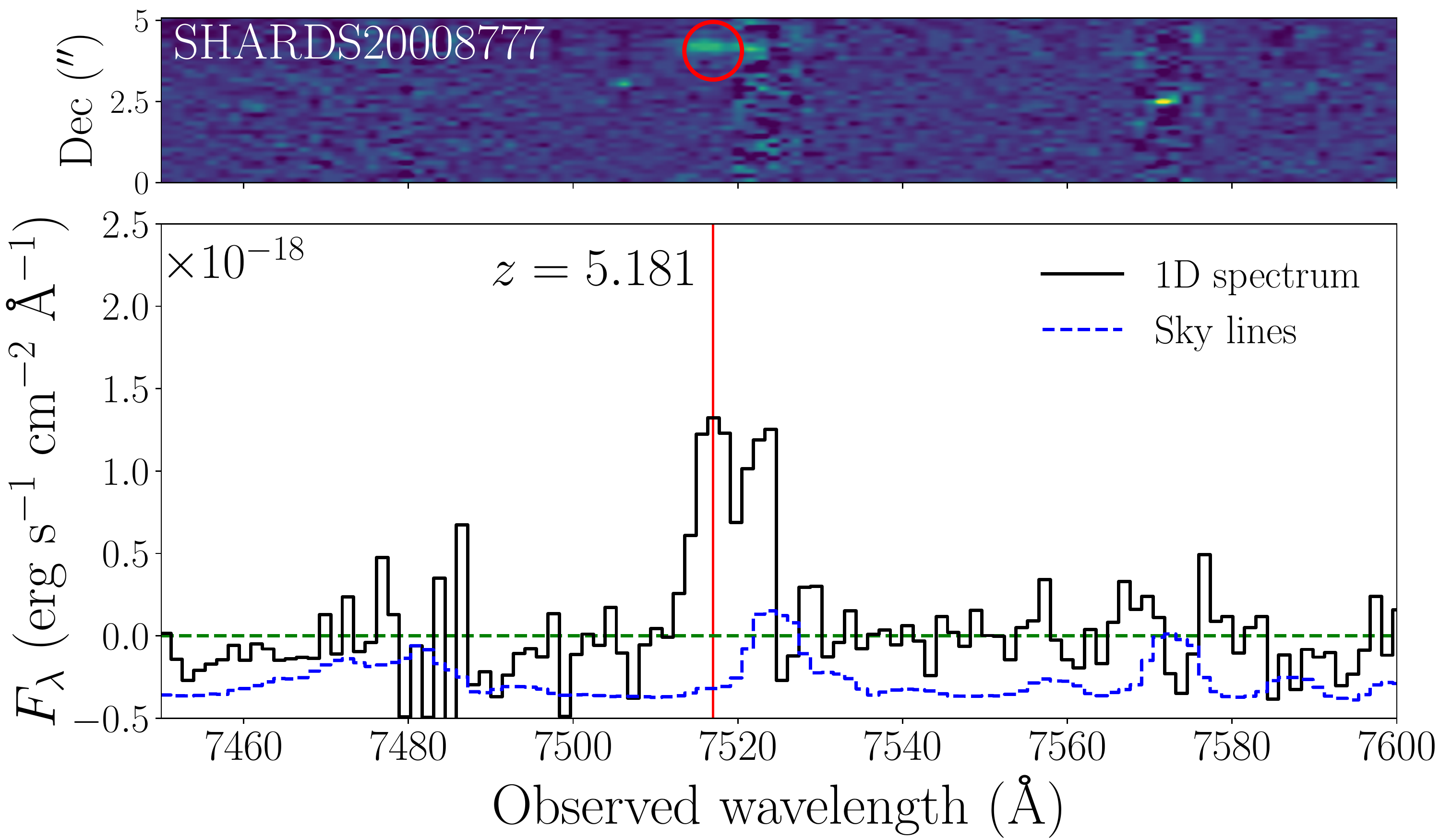} &
    \includegraphics[width=.425\textwidth]{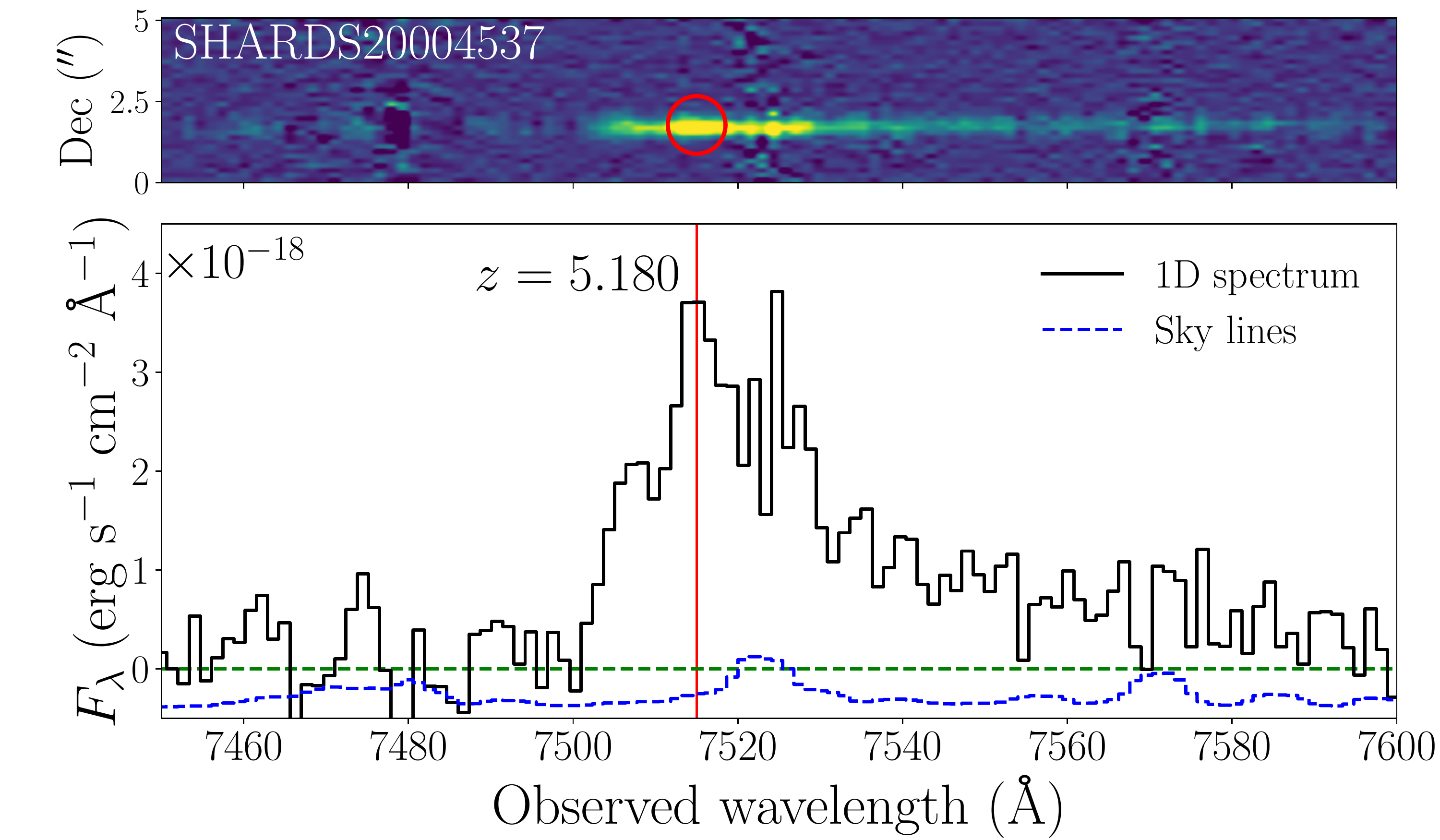} \\
    \includegraphics[width=.425\textwidth]{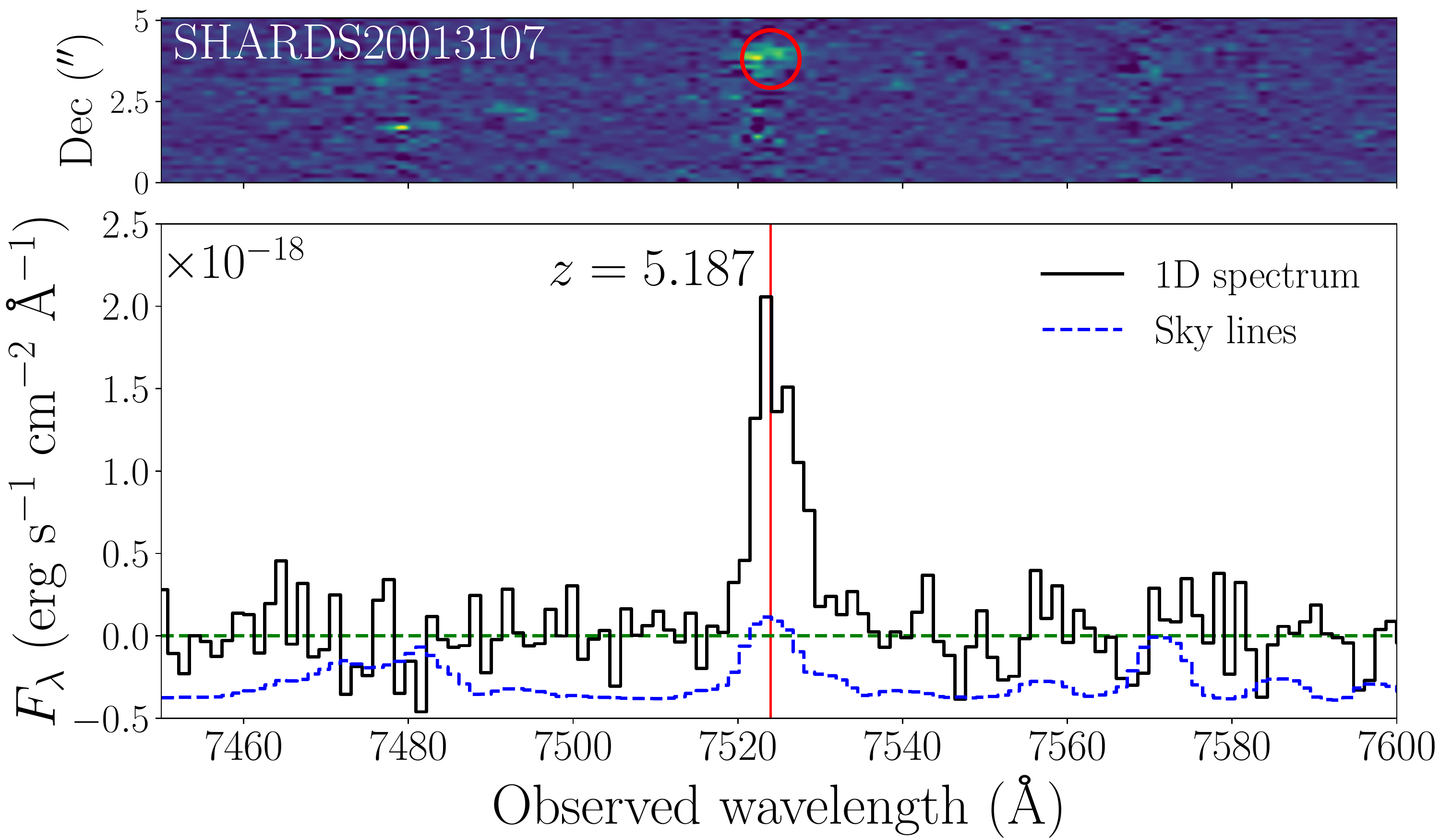} &
    \includegraphics[width=.425\textwidth]{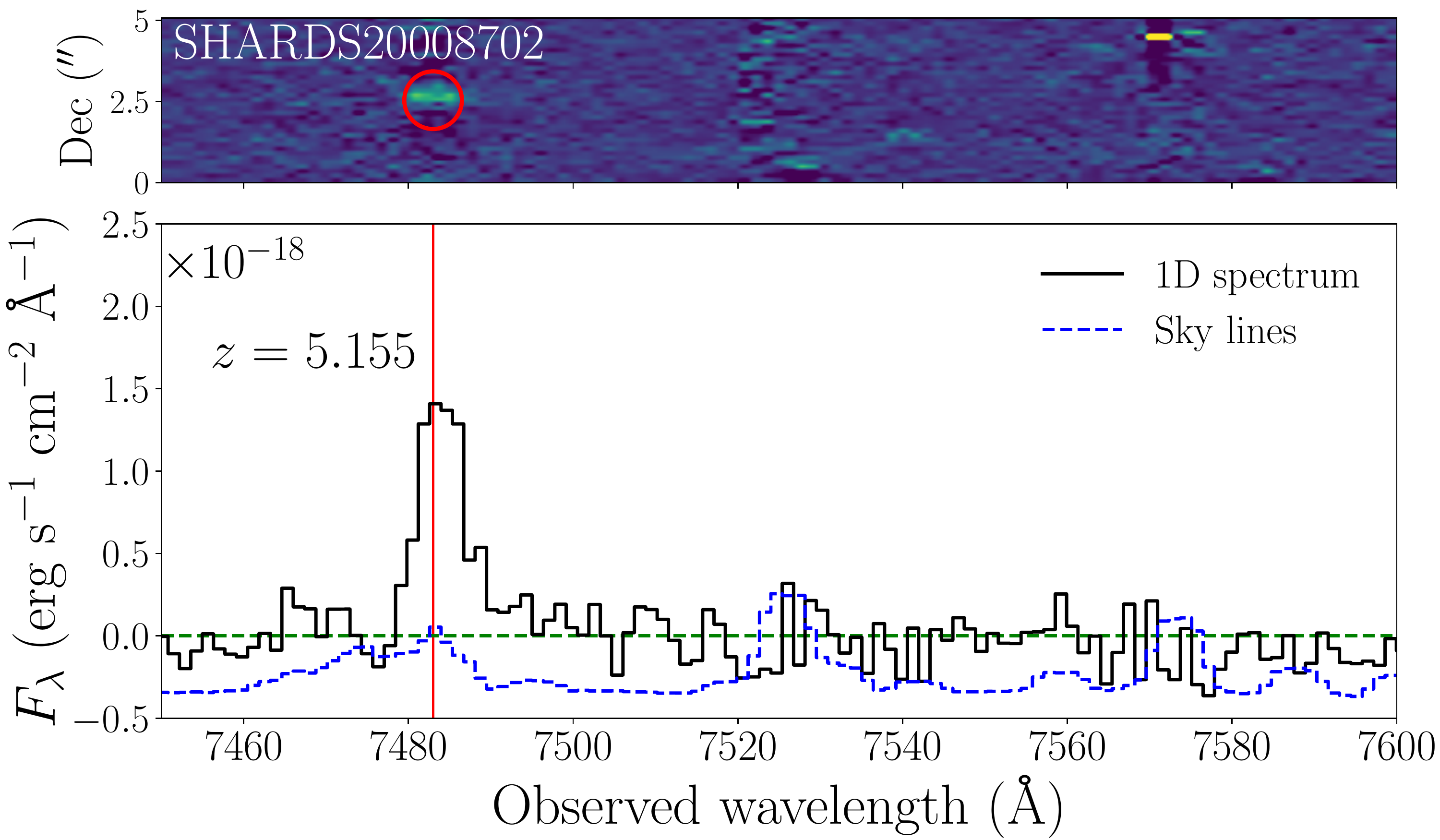}   \\
    \includegraphics[width=.425\textwidth]{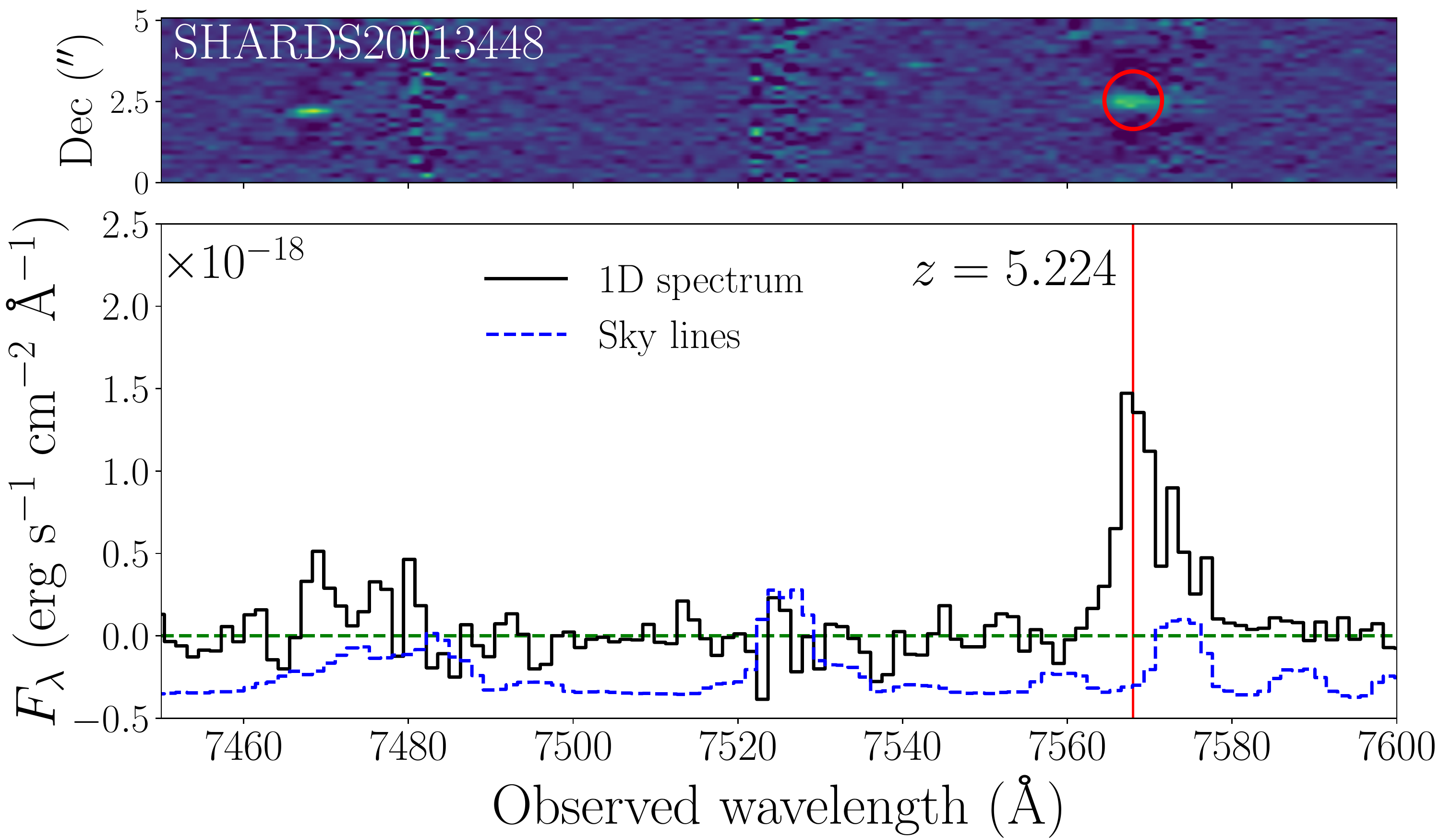} &
    \includegraphics[width=.425\textwidth]{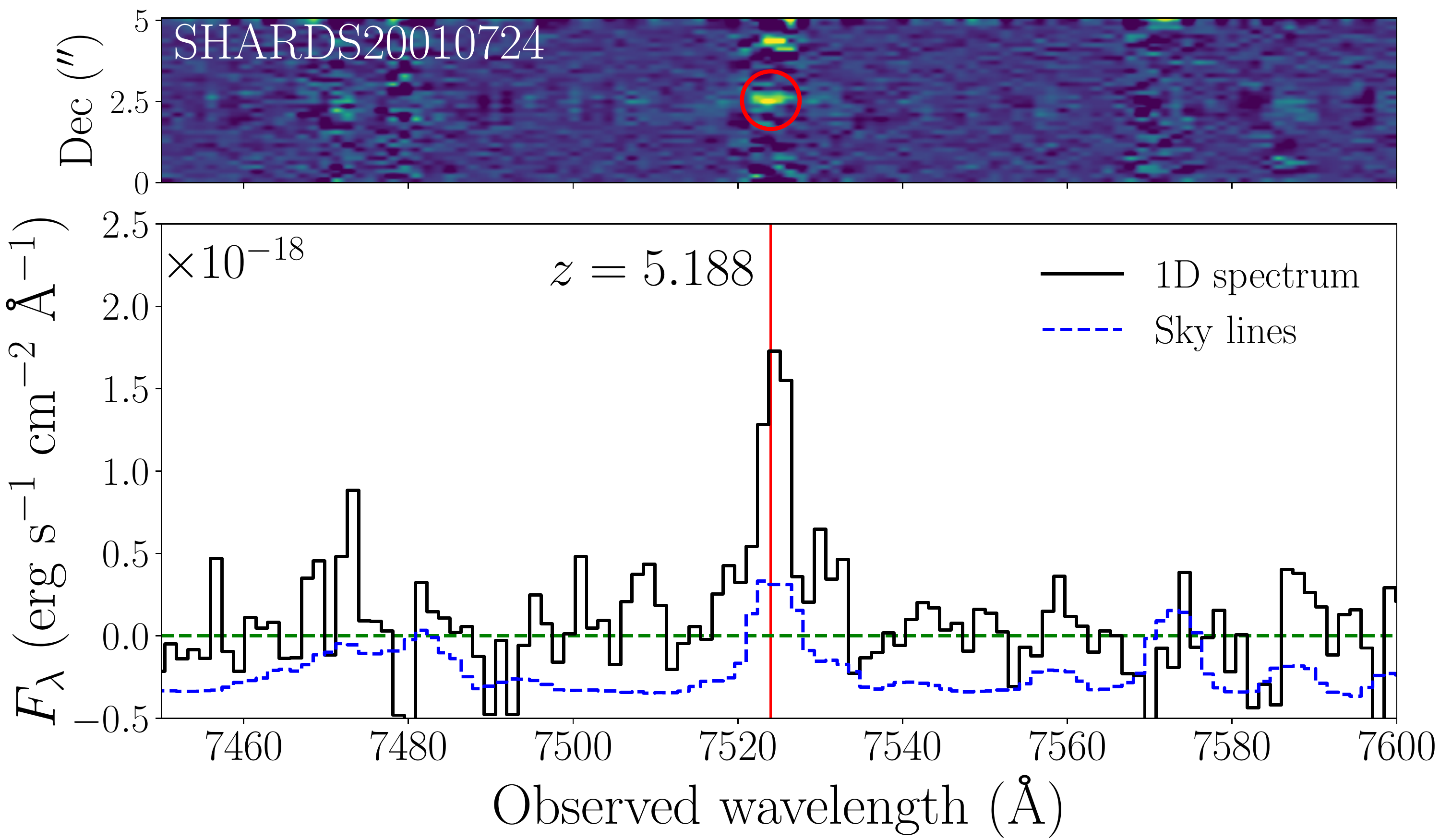}   \\
    \includegraphics[width=.425\textwidth]{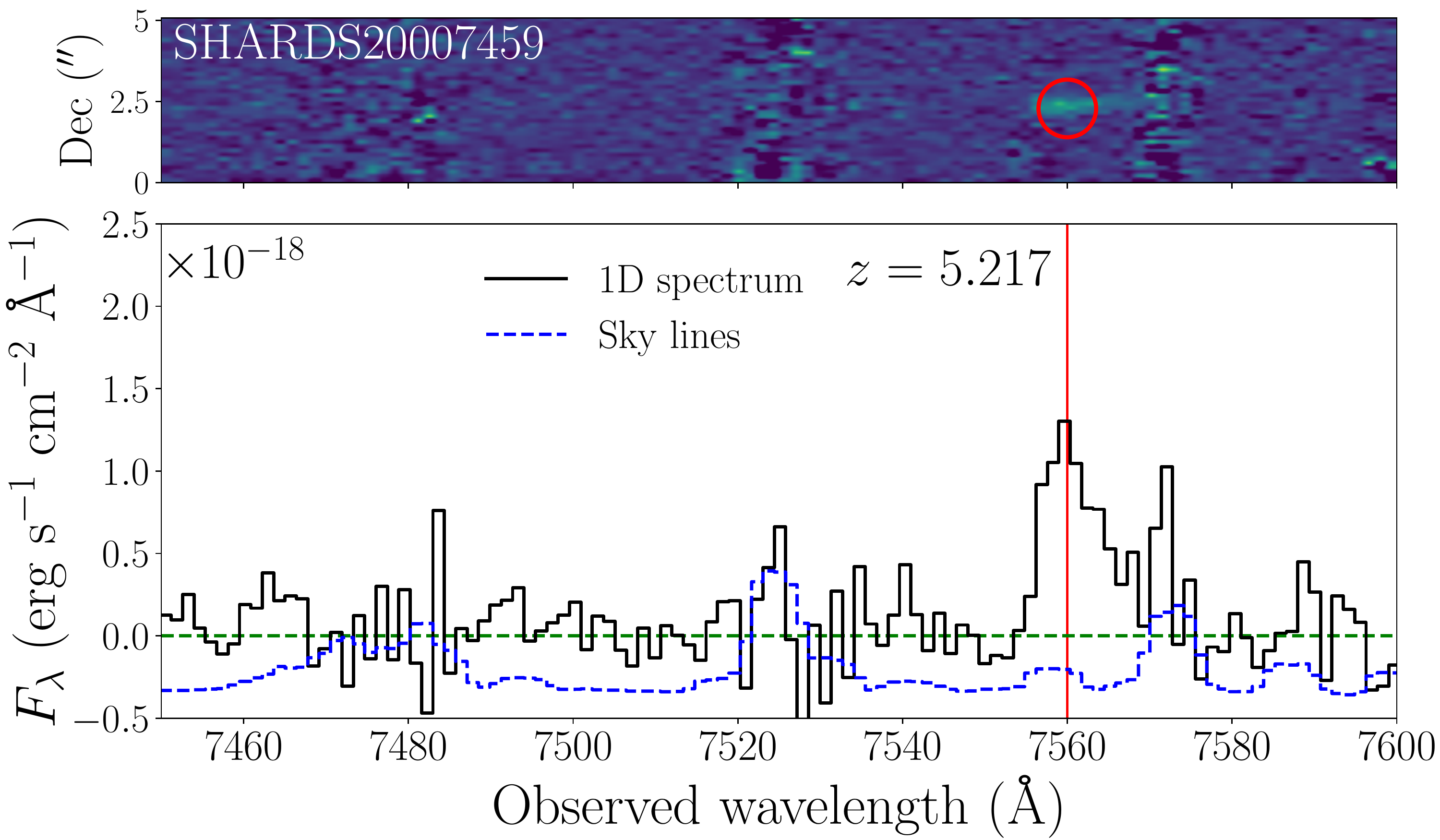} &
    \includegraphics[width=.425\textwidth]{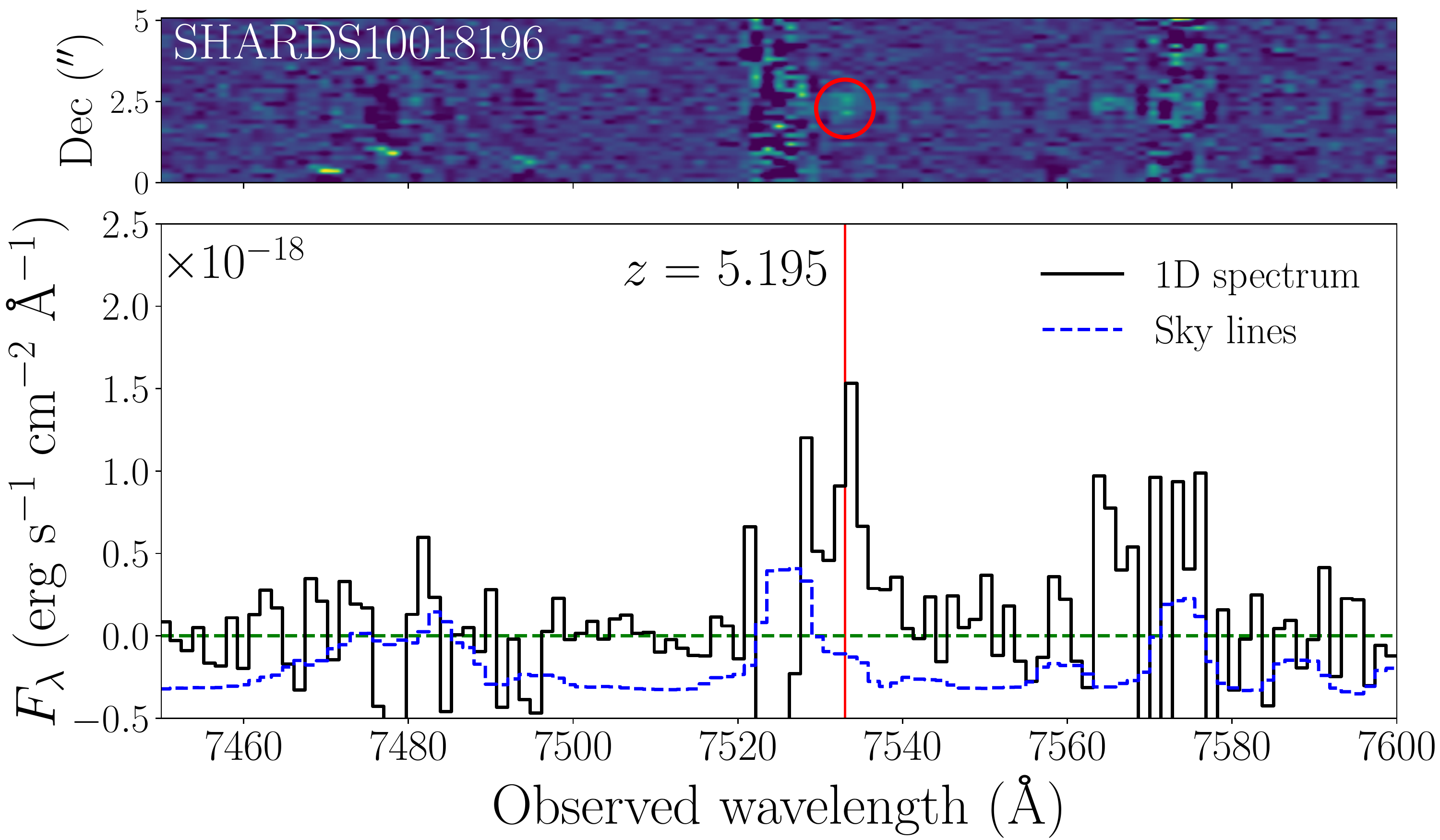}   \\
    \includegraphics[width=.425\textwidth]{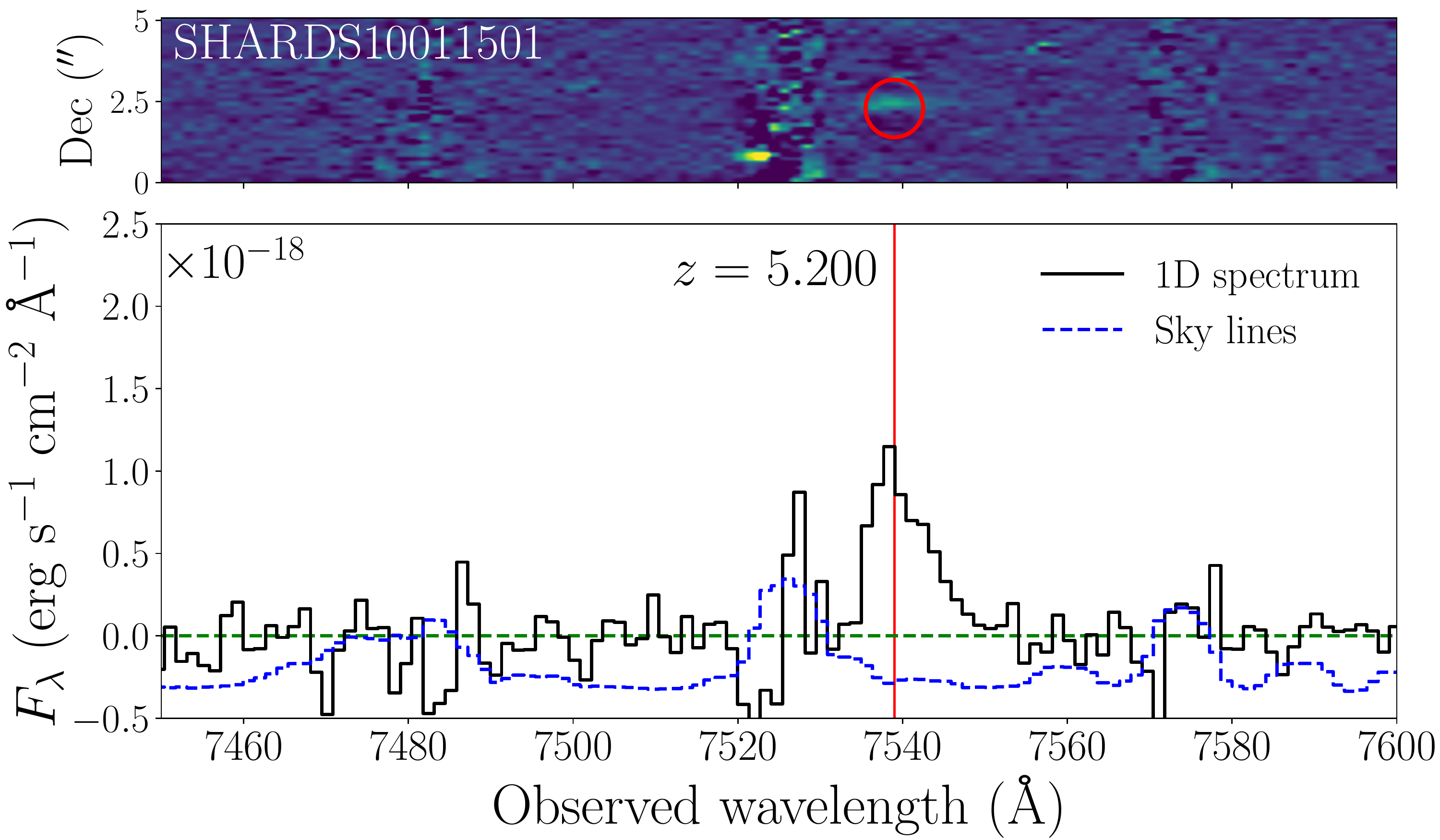} &
    \includegraphics[width=.425\textwidth]{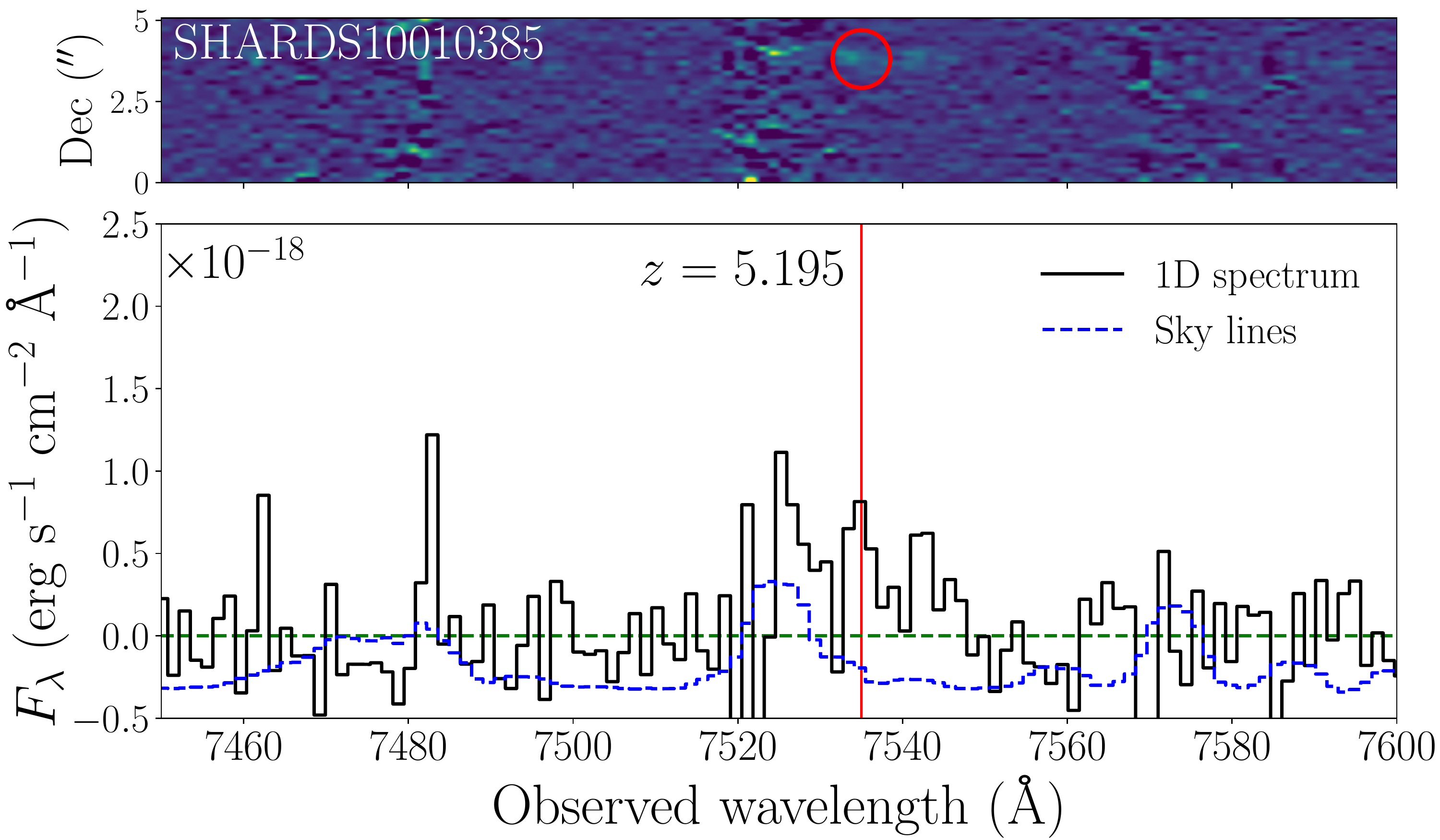}   \\
    
  \end{tabular}
  \caption{2D and 1D spectra of the sources classified like secure detection, grade A in Table~\ref{tab:detections}. The black lines are the collapsed one-dimensional spectra. The solid red line shows the peak of the Ly$\alpha$ emission line, with the corresponding redshift written besides. The blue spectra are the sky lines spectra scaled down so they do not interfere with the actual spectra of the sources. The horizontal green dashed line shows the zero flux level.}\label{fig:detectionA}
\end{figure*}

\begin{figure*}
   $\begin{array}{rl}
    \includegraphics[width=.425\textwidth]{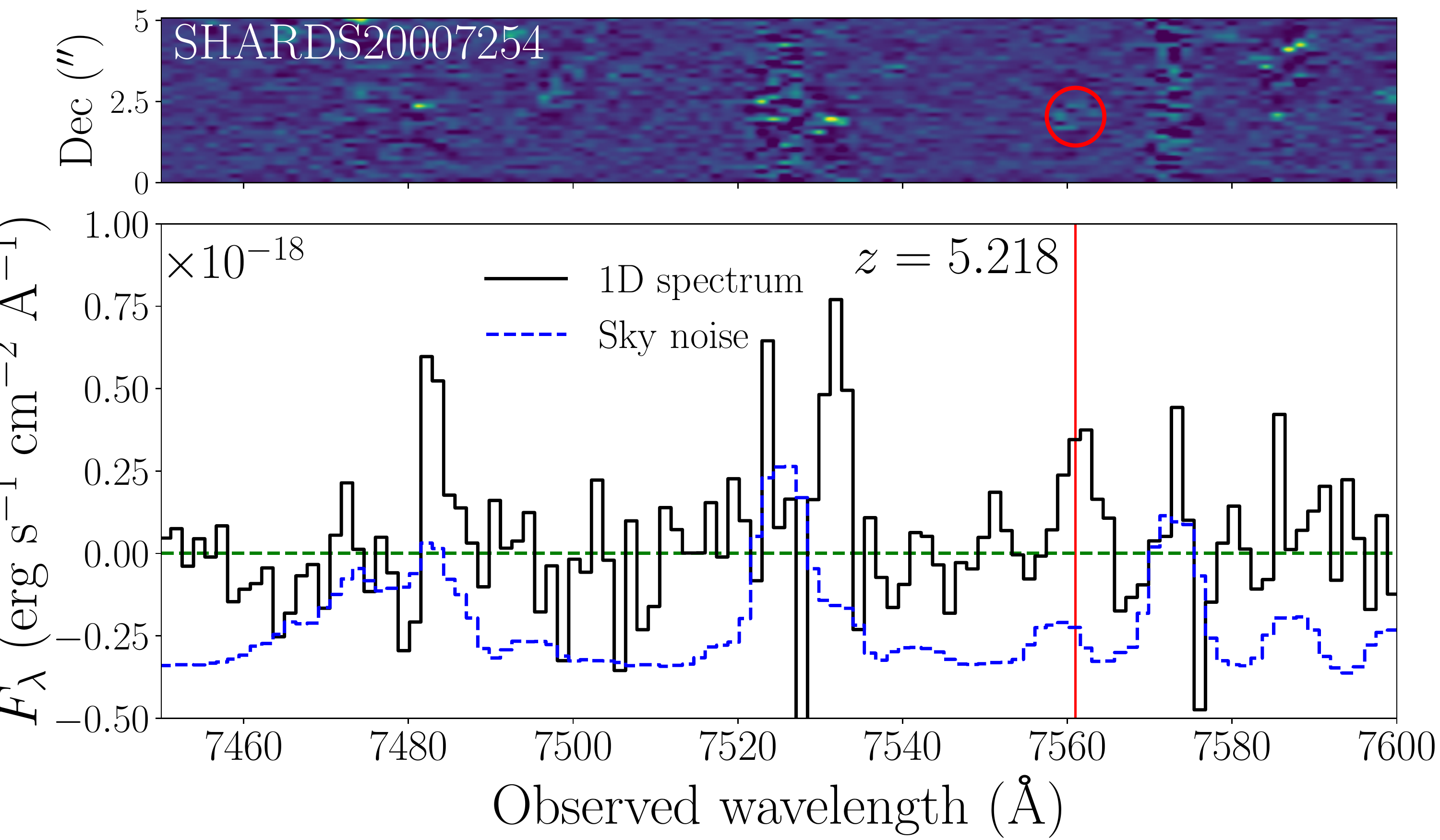} &
    \includegraphics[width=.425\textwidth]{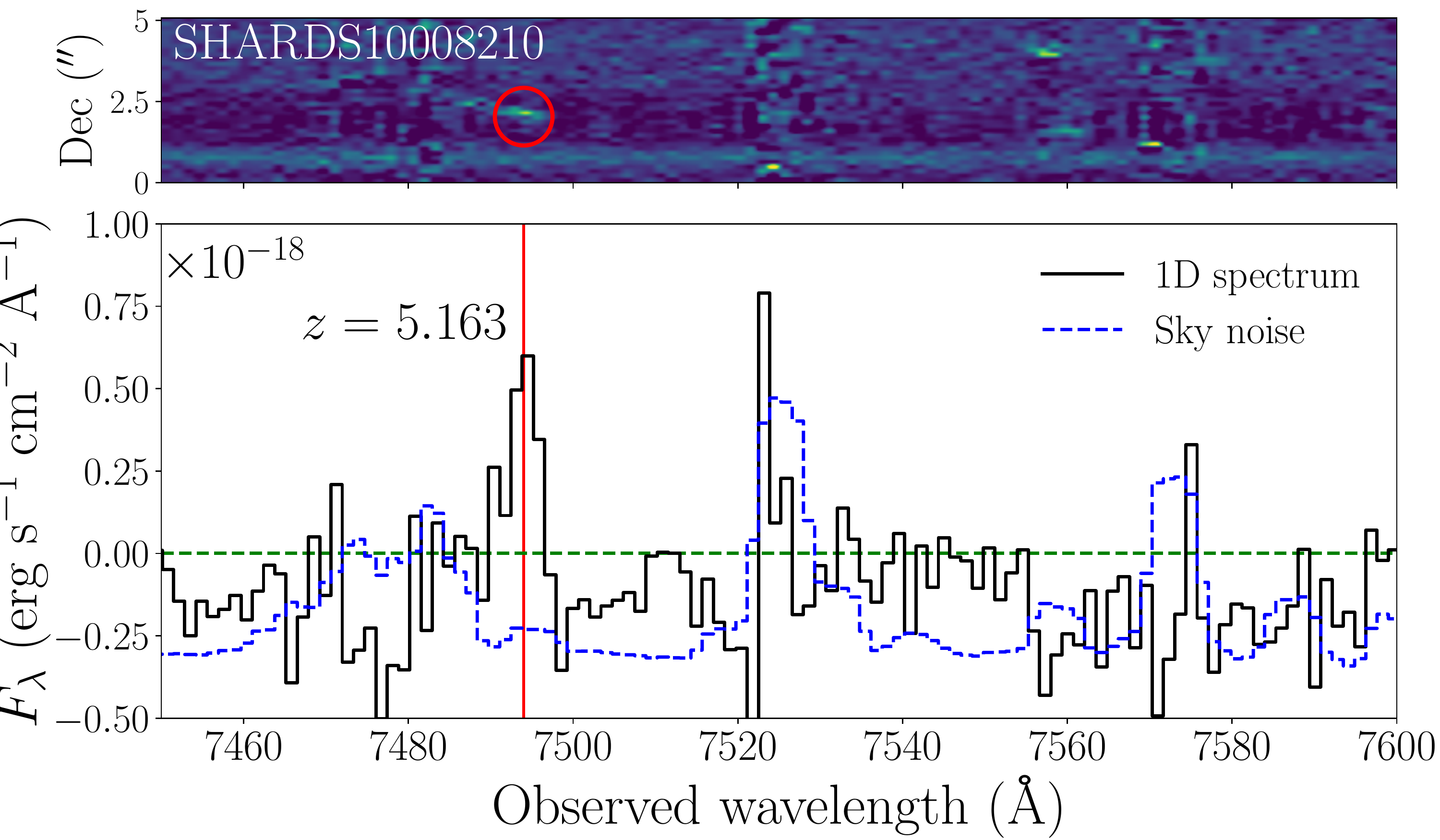} \\
    \multicolumn{2}{c}{\includegraphics[width=.425\textwidth]{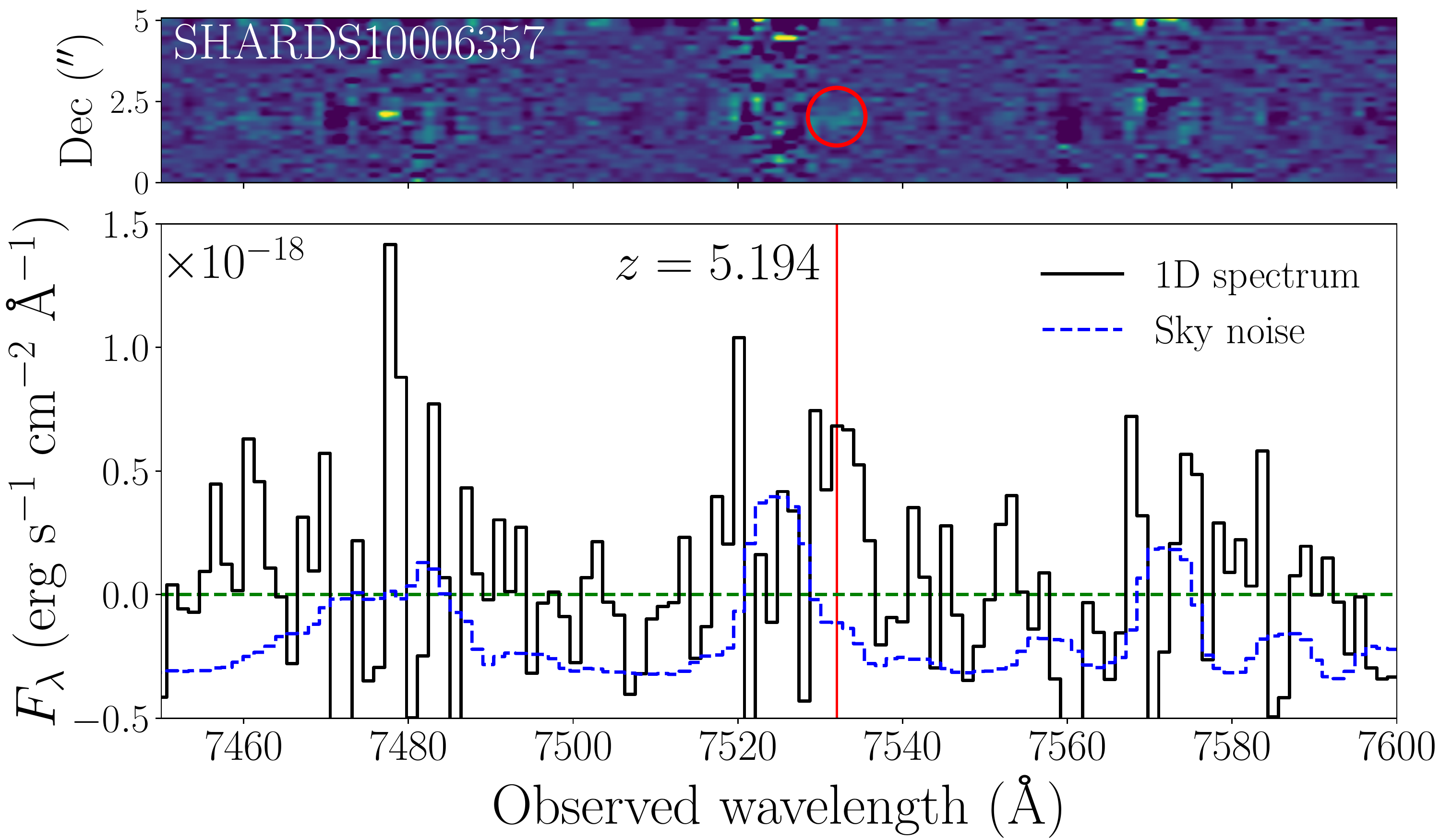}}   
    \end{array}$
  \caption{2D and 1D spectra of the sources classified like tentative detections, i.e. grade B in Table~\ref{tab:detections}. The meanings of all lines are the same as in Figure~\ref{fig:detectionA}. }\label{fig:detectionB}
\end{figure*}

\section{Results}
\subsection{Spectroscopic redshifts}
In total, we obtained spectroscopic redshifts for 13 out of 17 candidate protocluster members via the Ly$\alpha$ line. All 13 sources are physically associated with the protocluster. The spectra of the 13 cluster members, 10 with very good signal to noise (grade A) plus 3 additional sources with reasonably good signal to noise (grade B) are shown in Figure~\ref{fig:detectionA} and Figure~\ref{fig:detectionB}, respectively. The success rate of detections -- 13 out of 17 with spectroscopic redshift -- is 76\%. These sources show reliable line detection in the 2D spectra and signs of an asymmetric profile in the 1D spectra, resulting in reliable spectroscopic redshifts. The object IDs, coordinates, magnitudes, SHARDS photometric redshifts, obtained spectroscopic redshifts, source types (LAE or LBG) and quality flags (grades) are listed in Table~\ref{tab:detections}. The redshifts have been measured at the peak of the Ly$\alpha$ emission line simply using $z_{spec}=\frac{\lambda_{obs}}{1216 \mathring{A}}-1$. Several studies have demonstrated that most high-redshift galaxies have sufficiently high star formation rates to drive galactic scale winds. Signatures of the wind can then be seen both in the shape of Ly$\alpha$ line profile itself and in the red-wards velocity offset of the Ly$\alpha$ line of few hundreds of km/s with respect to the systemic redshift caused by the transit through the IGM's thick optical depth \citep{Shapley03,Steidel10,Hashimoto15,Hashimoto19}. 
However, the aim of this paper is not to probe the kinematic state of the ISM in these galaxies but to probe the existence of an overdensity at the redshift close to the SMG's one. Thus, the offset in velocity values does not affect the association with the SMG and we assumed that the peak marks the centre of the line.

The spectroscopic redshift  of these 13 sources ranges from $5.155\leq z\leq5.224$ ($\Delta z<0.069$). We compared the spectroscopic redshifts of these 13 confirmed members with the photometric redshifts derived from the SHARDS data \citep{Arrabal18}, see Figure~\ref{fig:scatter}. The resulting residuals $\delta_{z}=\left|z_{phot}-z_{spec} \right|$ range between $0.002\leq \delta_{z}\leq 0.055$, with $\bar{\delta_{z}}$=0.026, in close agreement with the $1:1$ relation between these two redshift measures. This analysis showed the very high reliability of the photometric redshifts derived by \citet{Arrabal18}, indispensable for the success of our spectroscopic observations.

We observed four sources from \citet{Walter12} to verify the reliability of our spectrocopic measurements. Three of them have been successfully recovered. Our measured spectroscopic redshifts are consistent with those from W12 (see Table~\ref{tab:detections} within the given uncertainties, accurate to the 3rd digit for both samples). The failure of the  non-detected source, an LBG galaxy with $m_{AB}>$~26.92 at the SHARDS filter F775w17, can be well explained by its very faint magnitude, more than $\sim$0.5~mag above the faintest sources with spectroscopic redshifts. The fact that some of our sources have been spectroscopically detected previously \citep{Walter12,Barger02} gives credibility to the spectra we have obtained. To summarise, our spectroscopic observations have increased the number of member galaxies, by 10 new confirmed members, from 13 to 23. 

\begin{figure}
  \includegraphics[width=\linewidth]{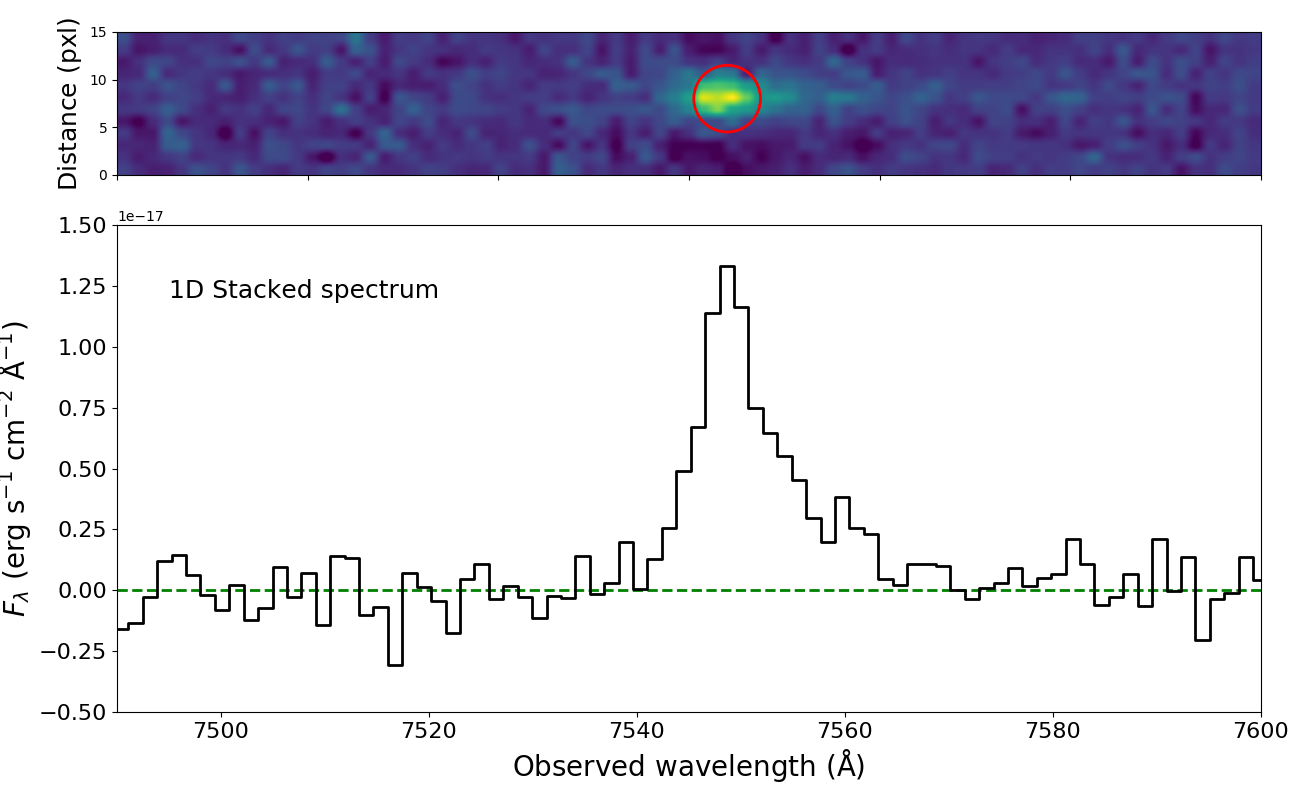}
  \caption{The stacked 2D and 1D spectrum of the 13 confirmed overdensity memb ers at $z=5.2$ with reliable detections.}
  \label{stack}
\end{figure}

\begin{figure}
  \includegraphics[width=\linewidth]{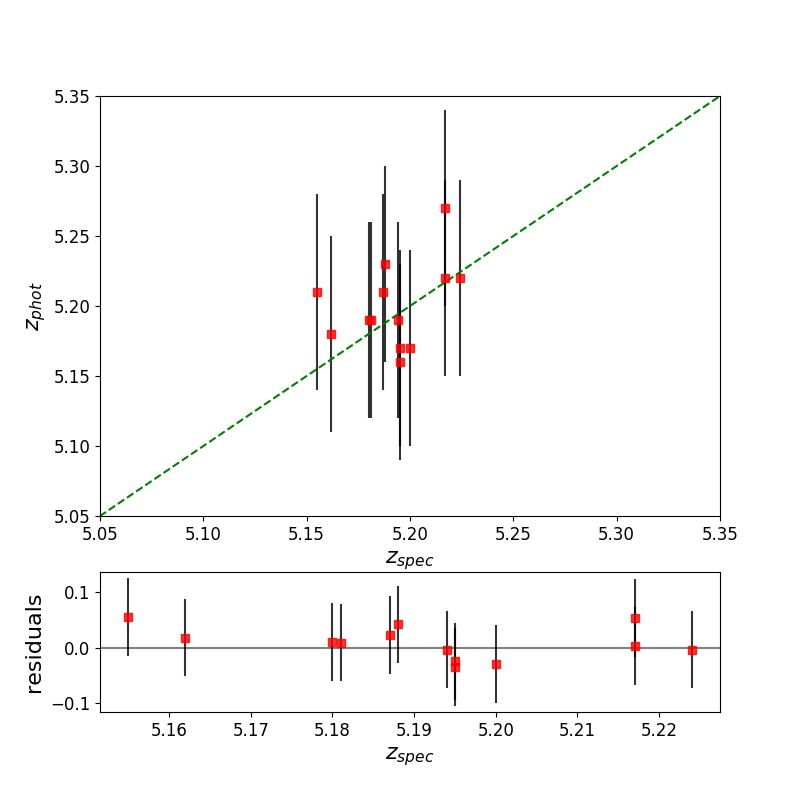}
  \caption{Comparison between the photometric and spectroscopic redshifts of the 13 LAEs classified with grade A and B.  The green, dashed line, is the 1:1 relation. In the bottom panel we show the residuals. This analysis shows that the photometric redshifts derived from the SHARDS dataset in GOODS-N in \citet{Arrabal18} are very accurate and thus very reliable for the search of overdensities.}
  \label{fig:scatter}
\end{figure}

A stacked spectrum was done by cutting all the confirmed spectra in a window of 100 pixels around the lambda-emission position, used as reference point and then summing up all frames obtained. The stacked spectrum is shown in Figure~\ref{stack}. It exhibits enhanced S/N as compared to the individual galaxy spectra. The asymmetry of the stacked line is noticeable. If the sky lines were contaminating the stacked spectrum we would expect a more symmetric line. Therefore, we have very high confidence in the 13 (both grade A and B) spectroscopically confirmed candidates that we have detected.

\subsection{Flux and Equivalent width measurements}
We measured the Ly$\alpha$ fluxes and rest frame equivalent widths (EW$_{0}$) manually, using IRAF, from the Ly$\alpha$ emission lines in the spectra, see Table ~\ref{tab:detections1}. This method guarantees a high degree of accuracy. Using the IRAF/SCOPY and IMSUM tasks, we located the position of the centroid of each emission line along the slit. Then, we defined an extraction window as $\pm$3 pix along the y-axis from the centroid. The total flux of a source is the sum along the six-contiguous spatial rows of this extracted spectrum. Then, using IRAF/SPLOT we marked two continuum points before and after the line. To determine the noise, we repeated the measurements five times for each source. The fluxes and the EW$_{0}$ for the 13 sources classified as A and B are shown in Table~\ref{tab:detections1}. As some sources are localised on top of sky lines, the errors in these cases were computed by adding to both the EW$_{0}$ and flux measurement errors of 25$\%$ of the mean value, a conservative approach. The values of EW$_{0}$ range between $75$~\AA\/ to $188~\mathring{A}$. The exception is the QSO whose EW$_{0}$ is above $300$~\AA\/. The high values of EW$_{0}$ are not unusual to find in a protocluster (see e.g. \citet{Steidel00,Dey16}). This indicates more efficient ionisation and more extreme ionising conditions of the nebular gas which are conducive to the escape of ionising photons \citep{Tang19}. In all cases beside one, the EWs obtained from the spectra are (much) larger than those derived from the SHARDS medium-band filters \citep[][given in parenthesis in Table~\ref{tab:detections1} of the present work]{Arrabal18}. This may indicate that only with spectroscopy a reliable EW could be determined \footnote{E.g. the case that the Ly$\alpha$ line falls outside the plateau of the filter response could be responsible that photometric measurements underestimate the EW.}. Using evolutionary synthesis models, \citet{Oti2010} show that once the equilibrium phase has been reached, the intrinsic Ly$\alpha$ equivalent widths should never be above $\sim100$~\AA\/. Almost all our sources lie above this theoretical limit.
Except for the QSO, our values of EW$_{0}$ are lower than the critical limit of Ly$\alpha$ rest frame equivalent width of $240$~\AA\/ obtainable from ionisation by a massive star population \citep{Charlot93}. These values are only consistent with very young (age < 10$^{7}$ years), almost coeval star formation episodes. 

\begin{table*}
\caption{Spectroscopically confirmed protocluster members in this work}
\label{tab:detections1}
\makebox[\linewidth]{

	\begin{tabular}{@{}lccccc}
    \hline
    \hline
\multirow{3}{*}{Source} & z$_{spec}$ & Flux$_{\mathrm{Ly}\alpha}$  & L$_{\mathrm{Ly}\alpha}$ & SFR$_{\mathrm{Ly}\alpha}$ & EW$_{0}$\\
    && $erg s^{-1}$ cm$^{-2}$ & $erg s^{-1}$ & M$_{\odot}\,yr^{-1}$ & \AA \\
    &&($\times 10^{-18}$) & ($\times 10^{42}$) & &   \\
    (1)&(2)&(3)&(4)&(5)&(6)\\
    \hline\hline
    SHARDS20008777&5.181 & 7.9$\pm$0.3 & 2.23$\pm$0.08 & 2.02$\pm$0.08 & 163$\pm$28 (110)\\
    SHARDS20004537&5.180  & 41.6$\pm$1.8 & 11.7$\pm$0.5 & 10.6$\pm$0.45& 315$\pm$70 (35)\\
    SHARDS20013107&5.187 & 13.1$\pm$0.2 & 3.70$\pm$0.06& 3.36$\pm$0.05& 155$\pm$61 (60) \\
    SHARDS20008702&5.155  & 10.4$\pm$3.0 & 2.91$\pm$0.84& 2.64$\pm$0.76& 149$\pm$61 (97) \\
    SHARDS20007254&5.218  & 2.4$\pm$0.1 & 0.70$\pm$0.03& 0.63$\pm$0.03& 120$\pm$25\\
    SHARDS20013448&5.224  & 9.1$\pm$2.3 & 2.62$\pm$0.66& 2.38$\pm$0.60&188$\pm$71 (89)\\
    SHARDS20010724&5.188  & 9.7$\pm$3.0 & 2.75$\pm$0.85& 2.50$\pm$0.77& 142$\pm$50 (111)   \\
    SHARDS20007459&5.217  & 9.4$\pm$0.4 & 2.70$\pm$0.11& 2.45$\pm$0.10& 166$\pm$16   \\
    SHARDS10018196&5.195  & 5.5$\pm$0.5 & 1.56$\pm$0.14& 1.42$\pm$0.13& 75$\pm$29 (78) \\
    SHARDS10008210&5.163  & 1.2$\pm$0.2 & 0.33$\pm$0.05& 0.30$\pm$0.05& 92$\pm$13 \\
    SHARDS10011501&5.200  & 8.4$\pm$0.1 & 2.40$\pm$0.01& 2.18$\pm$0.01 &156$\pm$18 (134)  \\
    SHARDS10010385&5.195  & 4.1$\pm$0.5 & 1.16$\pm$0.14& 1.05$\pm$0.13& 118$\pm$20 (26) \\
    SHARDS10006357&5.194  & 5.5$\pm$0.7 & 1.56$\pm$0.20& 1.42$\pm$0.18 & 129$\pm$52 (9)  \\
    \hline
    \hline
    \end{tabular}
}
\raggedright \justify
\textbf{Notes:}    
Column (1): SHARDS source name. Column (2): Spectroscopic redshift. Column (3): Ly$\alpha$ Flux. Column (4): Ly$\alpha$ Luminosity. Column (5): Star formation rate from Ly$\alpha$. Column (6): Rest-frame Ly$\alpha$ equivalent width. In parenthesis we give the equivalent width previously measured from the SHARDS medium-band filters \citep{Arrabal18}.  
\end{table*}

\subsection{Star formation rates}
To compute the star formation rates (SFR) we followed the standard Kennicutt calibration \citep{Kennicut1998}, assuming Case B recombination and a Salpeter Initial Mass Function (IMF) with a mass range of 0.1 to 100 $\rm M_{\odot}$, considering a $\frac{F_{\mathrm{Ly}\alpha}}{F_{\mathrm{H}\alpha}}$ flux ratio of 8.7 \citep{Brocklehurst71,Dopita03}.
The following conversion from luminosity to star formation rate was assumed:
\begin{eqnarray}
    \frac{SFR(H_{\alpha})}{M_{\odot}yr^{-1}}=\frac{7.9\times 10^{-42}}{8.7}\frac{L_{Ly\alpha,obs}}{erg s^{-1}}
\end{eqnarray}
where $L_{Ly\alpha,obs}$ is the observed luminosity computed as
\begin{eqnarray}
    L_{Ly\alpha,obs}=4\pi (d_{L})^{2}F_{Ly\alpha}
\end{eqnarray}
and $d_{L}$ is the luminosity distance calculated at each redshift.
This method follows a traditional and straightforward way of computing star formation rates. However, it should be noted that it is based on the assumption of a star formation episode producing stars at a constant rate during tens of Myr, until the birth and death of the most massive ionising stars reach equilibrium. The star formation rate for each source was directly derived from its observed Ly$\alpha$ luminosity. No attempts were made to correct for internal extinction as we do not have the possibility to determine the extinction for each object. Nonetheless, the (expected) extinction of LAEs at $z\sim 5.2$ is not  large, in general below A$_{V}=1$ \citep[e.g.,][]{Arrabal18,Bouwens15,Ouchi2009}.

\section{Characterisation of the $z=5.2$ overdensity}
In this section we discuss the properties of this protocluster at $z=5.2$. To carry out our analysis we will use all members spectroscopically confirmed by our work and the work by \citet{Walter12}. Thirteen sources come from our work and 10 from W12. We should note that three sources from W12 have been confirmed through our spectroscopic campaign and we use our measurements for the subsequent analysis. We listed all the members and their properties in Table~\ref{tab:members}. 
\begin{table*}
\small\addtolength{\tabcolsep}{-4pt}
\caption{All sample of the spectroscopically confirmed protocluster members}
\label{tab:members}
	\begin{tabular}{@{}lcccccccll}
    \hline
    \hline
Cluster member& Source name & R.A. & Dec. & $z_{spec}$ & M$_{\star}$ &z-range & Region &Reference & Comments\\
&& J2000.0 & J2000.0 & & (10$^{9}$ M$_{\odot}$)& & \it{Clump$^{\star}$}\\
(1) & (2) &. (3) & (4) & (5) & (6) & (7) & (8) & (9) & (10)\\
\hline
PCl$-$HDF850.1$-$01 & & 12:36:00.0& 62:12:26.1 & 5.199& 1.32$\pm$0.73 &$5.180\leq z\leq 5.208$& 4& W12\\
PCl$-$HDF850.1$-$02 & & 12:36:26.5 & 62:12:07.4& 5.200& 2.39$\pm$1.11&$5.180\leq z\leq 5.208$& 4& W12\\
PCl$-$HDF850.1$-$03 & & 12:36:37.5 & 62:12:36.0& 5.185& 1.52$\pm$0.94&$5.180\leq z\leq 5.208$& CR & W12\\
PCl$-$HDF850.1$-$04 & & 12:36:39.8& 62:09:49.1 & 5.187& &$5.180\leq z\leq 5.208$&2 & W12\\
PCl$-$HDF850.1$-$05 & SHARDS20004537& 12:36:47.96& 62:09:41.4&5.180&6.70$\pm$0.33&$5.180\leq z\leq 5.208$&2 &this work (B02, W12)&QSO\\
PCl$-$HDF850.1$-$06 & & 12:36:49.2& 62:15:38.6&  5.189&5.83$\pm$1.22&$5.180\leq z\leq 5.208$&1 & W12\\
PCl$-$HDF850.1$-$07 & SHARDS20013107& 12:36:49.79& 62:10:45.0& 5.187& 28.10$\pm$4.61&$5.180\leq z\leq 5.208$& 2 & this work\\

PCl$-$HDF850.1$-$08 & & 12:36:52.0& 62:12:25.8& 5.183&&$5.180\leq z\leq 5.208$& CR & W12&HDF850.1\\
PCl$-$HDF850.1$-$09 & & 12:36:55.4& 62:15:48.8 & 5.190&1.43$\pm$0.76&$5.180\leq z\leq 5.208$&1 & W12\\
PCl$-$HDF850.1$-$10 & & 12:36:55.5& 62:15:32.8 & 5.191 &3.74$\pm$1.98& $5.180\leq z\leq 5.208$&1 & W12\\
PCl$-$HDF850.1$-$11 & SHARDS20010724& 12:36:56.51& 62:13:13.6& 5.188 &6.43$\pm$4.39& $5.180\leq z\leq 5.208$& CR & this work\\
PCl$-$HDF850.1$-$12 & SHARDS20008777& 12:36:56.70& 62:09:30.5& 5.181&0.19$\pm$0.12& $5.180\leq z\leq 5.208$& 2& this work\\
PCl$-$HDF850.1$-$13 & SHARDS10010385& 12:36:58.43& 62:16:15.0& 5.195&1.97$\pm$1.16&$5.180\leq z\leq 5.208$& 1& this work\\
PCl$-$HDF850.1$-$14 & SHARDS20007459& 12:37:03.31& 62:13:31.5& 5.217&2.03$\pm$0.60& $5.217\leq z\leq5.224$&3 & this work (W12)\\
PCl$-$HDF850.1$-$15 & SHARDS20013448& 12:37:03.61& 62:11:58.5& 5.224&2.00$\pm$1.09& $5.217\leq z\leq5.224$&3 & this work\\
PCl$-$HDF850.1$-$16 & SHARDS10011501& 12:37:05.52& 62:16:01.3& 5.200&0.21$\pm$0.14& $5.180\leq z\leq 5.208$& 1& this work\\
PCl$-$HDF850.1$-$17 & & 12:37:09.9& 62:15:31.1 & 5.191&1.19$\pm$0.75 &$5.180\leq z\leq 5.208$&1 & W12\\
PCl$-$HDF850.1$-$18 & & 12:37:11.1& 62:16:38.6 & 5.208&10.20$\pm$2.02& $5.217\leq z\leq5.224$&1 & W12\\
PCl$-$HDF850.1$-$19 & SHARDS10018196& 12:37:12.48& 62:15:21.1& 5.195&0.25$\pm$0.80& $5.180\leq z\leq 5.208$& 1& this work\\
PCl$-$HDF850.1$-$20 & SHARDS20008702& 12:37:12.08& 62:10:54.1& 5.155&0.54$\pm$0.37& $5.155\leq z\leq 5.163$& - & this work\\
PCl$-$HDF850.1$-$21 & SHARDS20007254& 12:37:12.80& 62:11:32.0& 5.218&4.86$\pm$1.48& $5.217\leq z\leq5.224$& 3 &this work \\
PCl$-$HDF850.1$-$22 & SHARDS10008210& 12:37:14.50& 62:15:32.4& 5.163&7.07$\pm$4.02 &$5.155\leq z\leq 5.163$& 1& this work\\
PCl$-$HDF850.1$-$23 & SHARDS10006357& 12:37:15.63& 62:16:23.6& 5.194&3.29$\pm$1.15 &$5.180\leq z\leq 5.208$& 1& this work (W12)\\
\hline
\hline
\end{tabular}
\raggedright \justify
\textbf{Notes:} Column (1): Protocluster ID. Column (2): SHARDS source name. Column (3): J2000.0 right ascension of the targets. Column (4): J2000.0 declination of the targets. Column (5): Spectroscopic redshift. Column (6): Stellar masses. Column (7): Redshift range of the galaxy. Column (8): Reference {\it clump} (CR indicates the central region). Column (9): Proto$-$groups. Column (10): Additional comments.
\end{table*}

\subsection{Radio and far-infrared properties}
None of the 23 spectroscopially confirmed members (including HDF850.1) have a radio counterpart down to $\sim11\mu$Jy (5$\sigma$) in the extremely deep JVLA 1.4~GHz map of GOODS-N presented by \citep{Owen18}. This detection limit equals a star-formation rate of $\sim$1500-2000~M$_{\odot}$ per year, significantly higher than the star-formation rates derived for our sources from the Ly$\alpha$ line ranging between $\sim 0.3-3~$M$_{\odot}$ per year (see Table~\ref{tab:detections1}), and in the range of typical high-z dusty starbursts. We note that this structure would not have been found through the radio selection technique using the 1.4~GHz band to search for galaxy clusters in formation in the distant Universe \citep{Daddi17}. Furthermore, we searched the photometric public catalogue of 3306 `super-deblended' dusty galaxies obtained by \citet{Liu18} in GOODS-N. HDF850.1 seems to be marginally detected in the 1.4~GHz at S$_{1.4~GHz}\sim 11$~microJy \citep{Owen18,Liu18}\footnote{The flux is not reported in any solid detection list up to now.}. Beside HDF850.1, none of the cluster members are bright in the far-infrared/(sub)mm wavelength regime. In addition, no other significant molecular gas reservoir was revealed in this field at the redshift of HDF850.1 \citep{Riechers20}. 

 Beyond $z=5$ only a few systems with SMGs being the signpost of an galaxy overdensity confirmed through spectrocopy are known in the literature. One is the overdensity consisting of three LBGs physically related to the SMG AzTEC-3 \citep{Riechers10,Capak11,Riechers14,Pavesi18} and the other a pair of the dusty starburst galaxy CRLE at$z=5.667$ and the `normal' main-sequence galaxy HZ10 at $z=5.654$ \citep{Pavesi18}. In addition, photometric data indicate that both systems could be part of a  larger galaxy overdensity. In the case of the dusty starburst FLS3 at $z=6.3$ \citep{Riechers13}, a LBG overdensity cannot be completely excluded \citep{Laporte15}. To summarize, up to now no system --- dusty starburst surrounded by rest-frame UV bright galaxies --- is known with such a high number of spectroscopically identified members. The extent of our system at the sky is about $10^{\prime}\times10^{\prime}$ respectively, in physically scales about 4 $\times$ 4~Mpc, consistent with size predictions by \citet{Muldrew15} and \citet{Casey16}. HDF850.1 has a star-formation rate of $\sim850$ M$_{\odot}$ per year (W12). Whereas the sum of all 12 members (without the QSO) observed by us sum up to maximal $\sim40$ M$_{\odot}$ per year. Even applying an extinction correction the SFR of this protocluster is strongly dominated by the SMG. Within 1$^{\prime}$, we find three other rest-frame UV-bright galaxies. To summarize, we infer that the dusty starburst HDF850.1 could evolve into the brightest cluster galaxy and thus should be the center of this overdensity.

\begin{figure}
\centering
  \includegraphics[width=0.9\linewidth]{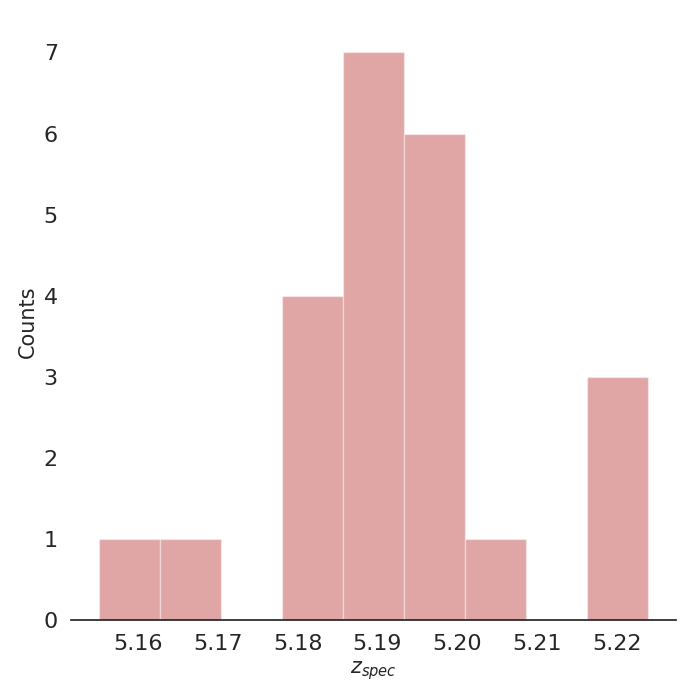}
  \caption{Redshift distribution of the 23 objects used in the friends-of-friends algorithm. For the protocluster galaxies in common with W12, we adopted our redshift determinations. Redshift bins are 0.01.}
  \label{fig:histo}
\end{figure}

\subsection{Clustering analysis}
In the standard $\Lambda$ Cold Dark Matter model ($\Lambda$CDM), the small halos form first, then merge to form larger ones. One model that may provide insights into the gravitational collapse of dark matter halos (DMH) is the spherical top-hat collapse model \citep{Gunn72,Gunn77,Peebles81}. This model assumes a highly idealized halo, with no interactions with the surrounding matter, characterised by its large overdensity $\Delta$ with respect to the background density. As the collapse starts, the overdensity $\Delta$ increases drastically and it is predicted to be $\Delta_{c}\sim 178$ times denser than the background. $\Delta_{c}$ indicates the critical density of a virialized halo. Some simulations suggest that this occurs at $\Delta_{c}\sim 200$, more or less independent of cosmology, and so a common mass estimator is M$_{200}$ which is approximately the virial mass if $\Omega_{M}=1$ \citep{White01}. 

According to N-body simulations the larger the scale the less the virialisation  \citep{Jang01, Hetznecker06, Davis10, Davis11}. With this in mind, the term `virialised' is not properly correct for high redshift protoclusters. Such structures typically are aligned along a filamentary structure and their members are unlikely to be governed by the virial theorem. 
However, even if the concept of virialisation is not the same referred to the massive bound DMHs by $z=0$, it is important to understand whether the distribution in both position and velocity space of our sources is consistent with the existence of any collapsing process at this redshift.
In the following we explore the spatial and velocity structures, the possibility of a ``core'' (virialised or currently virialising) and the existence of other substructures.

\subsection{Spatial and velocity fields}
\begin{figure}
\centering
\begin{tabular}{@{}cccc@{}}
  \includegraphics[width=1.1\linewidth]{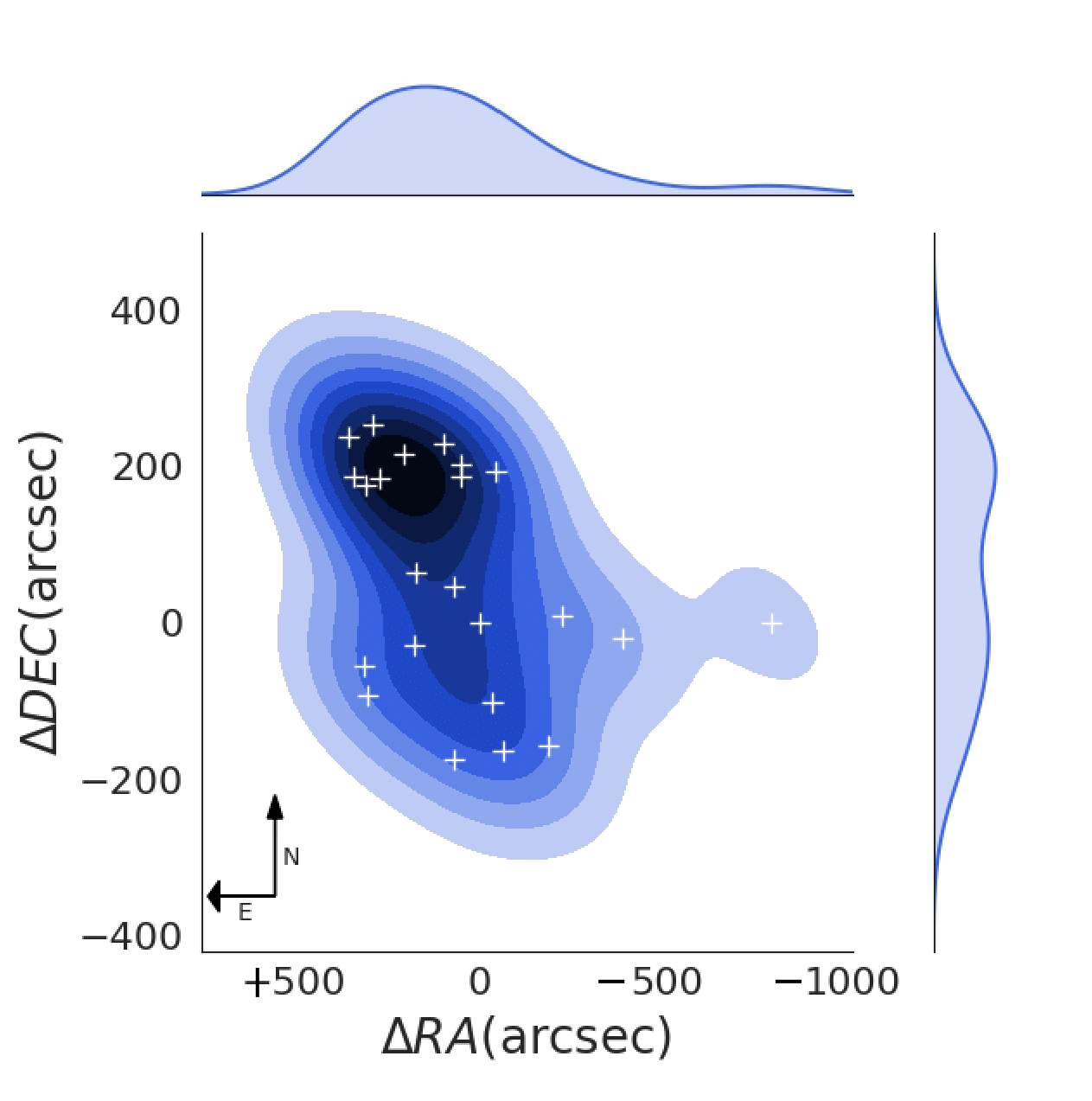}
  \end{tabular}
  \caption{Using kernel density estimation, we show the contour density plot in relative coordinates (offsets in RA and DEC). The origin (R.A.,DEC)=(0,0) is defined as the position of the SMG HDF850.1. The white crosses are the positions of the 23 sources. The marginal charts show the distribution of the 2 variables.}
  
  \label{density}
\end{figure}

The spectroscopic observations presented in this work confirmed 13 LAEs and LBGs \citep[three of them from][]{Walter12} belonging to the overdensity at redshift $z=5.2$. For a complete analysis we decided to include the 10 sources from \citet{Walter12}, having in total 23 sources physically related to the overdensity, including the dusty starburst HDF850.1. The redshift distribution of all 23 cluster members is shown in Figure~\ref{fig:histo}. The overall covered redshift is $5.155\leq z\leq 5.224$ ($\Delta z=0.07$). This range corresponds to a comoving radial distance of 34.8 Mpc (i.e. a angular size distance of $\sim 5.6$ Mpc). Assuming that the SMG lies at the center of the overdensity, the plot of contours of the objects surface density in Figure~\ref{density} illustrates that the protocluster lies predominantly in an elongated filamentary structure around the SMG along north-east to south-west direction. A density enhancement is located on the NE side while about half of the galaxies lie in regions that are underdense compared to that in NE suggesting that they are approaching the SMG from other directions along the filament. The histogram of redshift shows an evident peak at $z_{peak}\sim 5.19$ and two adjiacent peaks. The central bin includes 18 galaxies which range between $5.180\leq z\leq 5.208$ ($\Delta z=0.028$, $\Delta v_{LOS}\sim 1360 km/s$), the lower bin includes two galaxies with redshift $z=5.155$ and $z=5.163$ ($\Delta z=0.008$, $\Delta v_{LOS}\sim 390km/s$), and finally, three galaxies lie within the higher bin  $5.217\leq z\leq5.224$ ($\Delta z=0.007$, $\Delta v_{LOS}\sim 340 km/s$). The sources lying at the near edge of the redshift range (selected by our medium-band filter) could be 
`contamninants' (close to protocluster but not members of it) as well. However, we decide to keep these sources in the subsequent analysis because they could be part of a filamentary structure falling into the potential of the central high density region. As shown in Figure~\ref{density}, we identify two main regions: the NE region and the South region. In the NE region, we find a concentration of ten galaxies with redshifts higher than the SMG, nine within the central redshift bin ($5.189-5.208$) and one within the low-redshift  bin ($5.163$) which has an ambiguous spectroscopic redshift and could be an interloping galaxies that is not true member. The remaining galaxies are more widespread around the SMG (in the center of the plane) lying in a region streching  from SW to SE. Their redshifts fall within a broad range that includes all the three redshift bins ($5.155-5.224$), suggesting that they are associated with the protocluster but belong to physically distinct structures.

\subsection{Statistical properties in terms of stellar masses}

\begin{figure*}
\centering
\begin{tabular}{@{}cccc@{}}
  \includegraphics[width=0.8\linewidth]{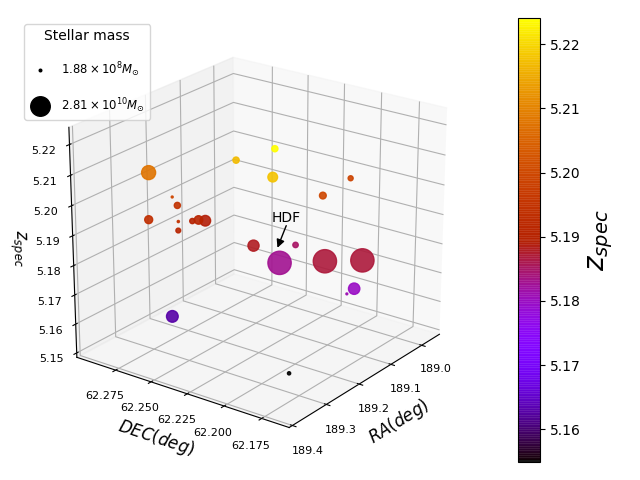}
  \end{tabular}
  \caption{3-dimensional plot of the distribution of galaxies in right ascension (x-axis), declination (y-axis) and redshift (z-axis). The circle size indicates the different stellar masses in the range $1.88\times 10^{8}M_{\odot}<M_{\star}<2.81\times 10^{10}M_{\odot}$. The color bar indicates the spectroscopic redshift.}
  
  \label{cube}
\end{figure*}

The stellar masses for the sources are estimated using the CIGALE code \citep[Code Investigating Galaxies Emission;][]{Burgarella05,Noll09,Boquien19}. This code has been developed to study the evolution of galaxies by comparing modelled galaxy spectral energy distributions (SEDs) to observed ones. We refer to \citet{Arrabal20} for details on the use of CIGALE to fit SEDs on the sample of 1558 high-z LAEs and LBGs selected from the SHARDS galaxies in the range $3.4<z<6.8$. We provide the stellar masses for 21 out of 23 members which are part of our LAE/LBG sample. In case of HDF850.1 (\#8) and the galaxy (\#4) no stellar masses can be provided (the latter one is not part of our sample). Stellar mass  estimates are shown in Table~\ref{tab:members}. These values lie in the range $(0.018-2.81)\times 10^{10}$M$_{\odot}$ with a median M$_{\star}$ of $(2.03\pm1.67)\times10^{9}$M$_{\odot}$. In Figure~\ref{cube} we show the 3-dimensional distribution of all 23 galaxies according with their stellar masses (indicated with different sizes) and redshifts (indicated with the color bar). We include the galaxies \#8 and \#4, for which we do not have the stellar mass estimates, considering for them the highest value of M$_{\star}$ in our list. It is clear that the LAEs/LBGs fall into several possible velocity groupings along the line of sight. A number of galaxies with redder colors (lying at higher redshift than HDF850.1) and low-intermediate masses are grouped on one side (NE) of the SMG galaxy. This concentration of galaxies in the NE region, already seen in the density plot in Figure~\ref{density}, could be an indication of a population of 
`proto-red' sequence galaxies'. The presence of a passive sequence, similar to that in low redshift clusters, has been both demonstrated in observations in protocluster at $z\sim 2-3$ \citep{Lemaux14,Wang16,Diener13,Cucciati14,Strazzullo16} and predicted in simulations \citep{Contini16}. Indeed, the {\it red sequence method} is a powerful tool for finding clusters by looking for overdensities of galaxies that form a red sequence \citep{Kurk04,Kodama07,Zirm08}. Several galaxies with violet-blues colors appear located on the opposite side of the SMG. Additionally, there is a number of galaxies a bit further respect to the SMG that might represent a filamentary distribution within the overdensity. 

\begin{figure}
\centering
\begin{tabular}{@{}cccc@{}}
  \includegraphics[width=\linewidth]{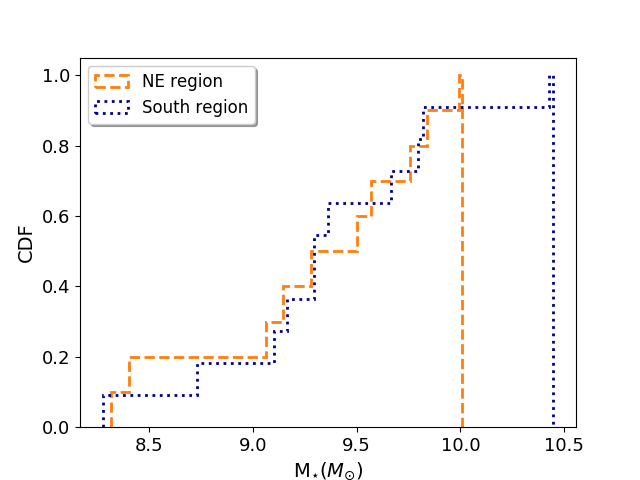}
  \end{tabular}
  \caption{Cumulative distribution function of the stellar masses in the NE region (orange dashed line) and in the South region (dotted blue line).}
  
\label{fig:mass}
\end{figure}

From a statistical analysis of the stellar masses, we try to understand if the galaxies in the NE region could be the sign of a more evolved population of more massive galaxies. In Figure~\ref{fig:mass} we plot the cumulative distribution functions (CDFs) of the stellar masses in the two main regions in which the protocluster is separated. Most of the galaxies, $\sim$70-80\%, range from  10$^{9}$M$_{\odot}$ to 10$^{10}$M$_{\odot}$ while only 20\% have a stellar mass less than 10$^{8}$M$_{\odot}$ regardless of their location in the protocluster. Using a two-sample KS test, we also investigate whether the mass measurements of galaxies in NE and South regions are consistent with being drawn from an incomplete but uniformly distributed sample on the FoV. We find a $p$ value of 99.9\%, thus we cannot reject the hypothesis that the distributions of the two samples are the same. Finally, we do not find any strong evidence of an enhancement of the stellar mass density in the NE side, compared to the South one where galaxies are more dispersed, but we can not avoid to note the peculiarity of this substructure. Indeed, the galaxies in the NE region appear to be assembled faster with a star formation activity still on-going.

\subsection{Evidence of substructures}
To assess the existence of any grouping or clustering which can characterize underlying substructures in the density field of our protocluster we use a Friend-of-Friend (FoF) algorithm which follows the method developed both to search for groups and clusters in a magnitude limited survey \citep{Huchra82,Eke2004,Berlind06} and in cosmological simulations \citep{Einasto84,Davis85,Lacey94,Jenkins01,Gottl07}. This algorithm select groups which correspond as closely as possible to real 3D groups, at least in a statistical sense. An iterative procedure is applied, according to which two galaxies belong to the same system if their projected separation and their line of sight ($los$) velocity dispersion are less than a fixed threshold. 
It should be noted that the shifting of the Ly$\alpha$ line center due to the photon strong resonant scattering from neutral hydrogen in the interstellar medium (ISM), produces large uncertainties in the measurement of the velocity dispersions. However, when estimating the overdensity structure, we assume an isotropic velocity distribution in 3D space and the differences in redshift are caused by their $los$ component velocities within the redshift space. With this assumption the component $\sigma_{\mathrm{los}}$ of the velocity dispersion is $\frac{\sigma}{\sqrt(3)}$.

We first adopt as linking parameters a projected mutual distance $D_{L}$ = 0.5~Mpc, that is almost two times the radius R$_{200}$ of a group, and a $los$ velocity dispersion, V$_L$, fixed at 500~km/s rest frame to avoid the interlopers. We chose these linking lengths to explore the systems within the protocluster that already constitute a real interacting group. For each galaxy we searched for its first neighbour and, by a recursive procedure, we added neighbours of neighbours until no more were found. We identify in the NE region two gravitationally bounded systems: a group of three galaxies and another of six galaxies. We define these systems as ``trial'' group. To study the internal dynamics of the two systems in the NE region with $N_{gal}\geq3$ (from now on group$-$I and group$-$II) and eventually asses the presence of a ``core'' in the protocluster, we required a more accurate membership determination. Indeed, the accidental inclusion of interlopers is one of the major problems in identifying groups in redshift space. For these two groups  we used the well-defined statistical approach by \citet{Beers90} and we computed the position of the geometrical centre in each group and the gapper velocity dispersion. The Gapper algorithm is a robust statistical estimator recommended over the canonical rms standard deviation as this algorithm is insensitive to outliers and thus reproduces more accurately the true dispersion of the systems with sizes $<10$ \citep{Beers90}. The velocity dispersion is given by:
\begin{eqnarray}
\centering
    \sigma_{gapper}=\frac{\sqrt(n)}{n(n-1)}\sum_{i=1}^{n-1}w_{i}g_{i}
\end{eqnarray}
where $w_{i}=i(n-i)$ and $g_{i}=x_{i+1}-x_{i}$ is the velocity dispersion. Then, we restricted the membership to galaxies within $\pm \sigma_{\mathrm{los}}$ from the median group redshift and located within a projected distance of $\pm R_{200}$ from the geometrical centre. Assuming that these systems are collapsing (approximately virialised), R$_{200}$ is an approximation of the radius which defines a sphere with the mean interior density 200 times the mean matter density at that epoch ($\langle \rho\rangle  (1+z)^{3}$). To obtain an estimate of the radius R$_{200}$, we follow the same procedure as \citet{Chanchaiworawit19}. We assume an NFW halo model \citep{Navarro96,Navarro97} and with a given $\sigma_{los}$, the radius R$_{200}$ is expressed as
\begin{eqnarray}
R_{200}=3\sigma_{los}\Bigg(\frac{1}{560\pi G(1+z)^{3}\langle \rho \rangle}\Bigg)^{\frac{1}{2}}
\label{radius}
\end{eqnarray}
where G is 4.9$\times$10$^{-9}$(Km/s)$^{2}$Mpc$^{-1}$M$_{\odot}$. We iterated the process till when the last two iterations have identical output. We find that only a pair of galaxies in the group$-$I are bound within R$_{200}$=104 kpc. 
The pair of galaxies in the group$-$I have spectroscopic redshifts of 5.19 and 5.191 and a relative velocity of 300km/s.. From the value of $R_{200}$ we computed the associated dynamical masses $M_{200}$ using the equation 
\begin{eqnarray}
   \frac{3M_{200}}{4\pi R_{200}^{3}}=200\langle \rho \rangle (1+z)^{3}
    \label{equ:mass}
\end{eqnarray}
The derived mass is $M_{200}\sim$9.4$\times$ 10$^{12}$M$_{\odot}$.
The high mass estimate can be explained as a natural bias because clumps in high density regions collapse earlier and accrete faster their mass. These values are rare but not impossible to find at z>5 (e.g. \citet{Chanchaiworawit19}). In any case, under the assumption of virialisation and spherical symmetry for the system, this mass has to be considered an upper limit. 

These results confirm a segregation of galaxies in the protocluster. A more evolved halo is located in the densest NE region. All galaxies in this region range within a projected distance of $<670$ kpc and a probable case of an `off-centre' core is present. A pair-like structures are also found by \citep{Toshikawa14,Topping16,Toshikawa20}. These authors suggests that large-scale galaxy/group assembly start by $z\geq4$ with primordial satellite components that appear in parallel with the formation of central protoclusters. The NE region might be considered as the very massive halo, where galaxies are dynamically bound and are falling into the potential of the central high density region. We define the NE region as the `clump1' of the overdensity.

An analysis of the velocity distribution of the southern sources have shown that a fraction of the galaxies have relative low velocities even if they still did not enter in a common halo. Thus, they can be part of dynamically young systems in the process of merging. With this purpose, we investigated these galaxies by considering different and increasing values of $D_{L}$ (550-600-700-800 kpc) and we run again the FoF. We identify four substructures. The first is located in the central part of the overdensity structure. It hosts the SMG that is separated by $<$500 kpc in physical distance from the closer galaxies and their relative $\sigma_{los}\sim$300km/s. Since the SMG should be the proto-BCG of the overdensity, we find that its formation history is possibly the result of multiple high-z progenitors which could assemble into a single BCG at low-z, in agreement with simulations \citep{Ragone18}. We define it as `Central region'. The small group of three galaxies associated to the QSO CXOHDFNJ123647.9$+$620941 \citep{Barger02,Barger03}, reside on the periphery of the structure. They have a physical distance from the centroid of the group within $\sim$300~kpc and a velocity dispersion $\sigma_{los}$=281km/s. We define it as `clump2'. Another substructure in the SE region is made up by three galaxies --- in the highest redshift $5.217\leq z\leq 5.224$ ---, with a physical distance within  $\sim$250~kpc from the center of the group and $\sigma_{los}$=400km/s. We name it `clump3'. Finally two galaxies on the SW side with a physical distance within $\sim$100~kpc and $\sigma_{los}\sim$50~km/s is defined as `clump4'. The galaxies in the last two groups appear clustered in redshift as well as spatially. However, their small number can not allow us to draw any definitive conclusions. They can be attributed to some filaments along the line of sight, providing further evidence that the structure is dynamically young.

\begin{figure*}
\centering
\begin{tabular}{@{}cccc@{}}
  \includegraphics[width=0.7\linewidth]{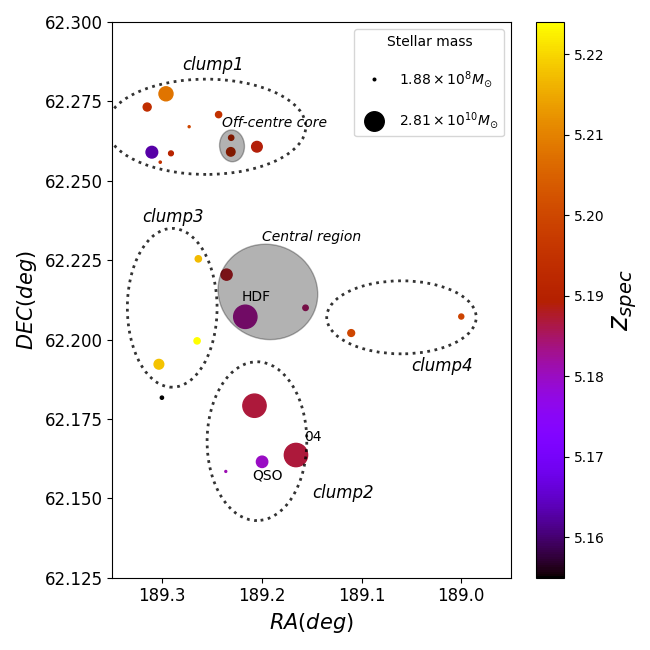}
  \end{tabular}
  \caption{Position  of the 23 sources used in the FoF analysis in right ascension and declination in degree. The sizes of the circles are in correspondence of different stellar masses (for the SMG and the galaxies \#4 we considered the highest value of M$_{}\star$), while the color bar indicate the spectroscopic redshift. The dotted black ellipses enclose LAEs in the NE Clump1, south central Clump2, SE Clump3 and SW Clump4. The shaded gray areas indicate the pair-like system in the NE region and the central region of the overdensity around the SMG.}
  
  \label{plane}
\end{figure*}

The separation between the Central Region and the mean redshift of the four clumps is within $\Delta z<0.04$ and is equal to 484~kpc, 170~kpc, 2.75~Mpc, 1.2~Mpc in physical scale, for clump1, clump2, clump3 and clump4, respectively. We show these clumps in the 2-dimensional plot in Figure~\ref{plane}. The sky distribution in the 3D and in the 2D plot evidences the presence of a few galaxies that do not belong to any substructure. They are probably located at the outskirts, suggesting the existence of a larger filamentary structure.

\subsection{Phase-space diagram}
\begin{figure}
\centering
\begin{tabular}{@{}cccc@{}}
  \includegraphics[width=1.1\linewidth]{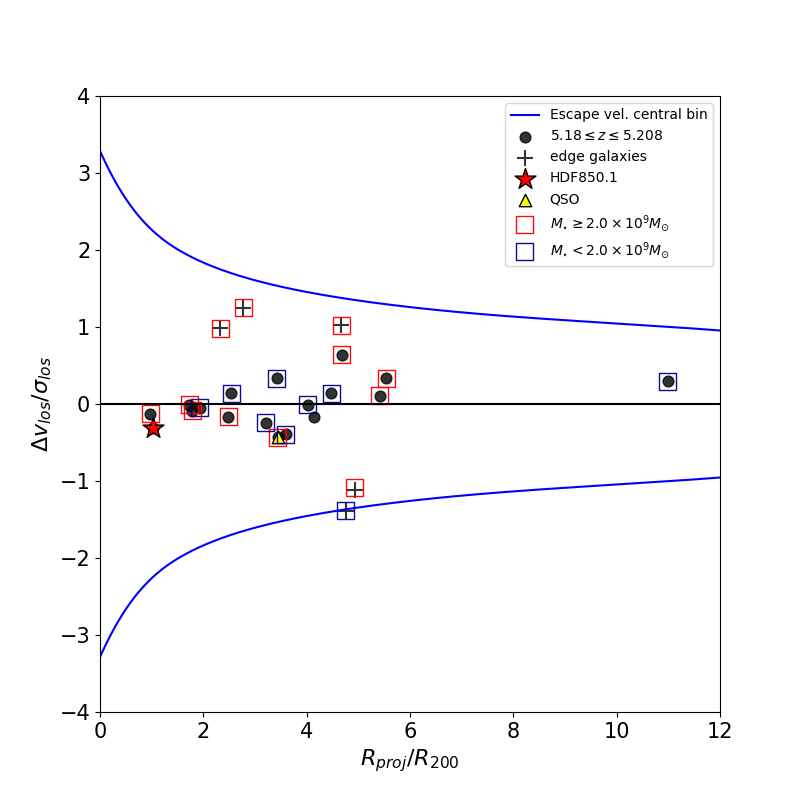}
  \end{tabular}
  \caption{The observed phase-space diagram for the  spectroscopically confirmed member galaxies in PCl-HDF850.1. The black circles are the 18 galaxies in the central bin $5.18\leq z\leq 5.208$. The black crosses are the galaxies in the lower and higher redshift bin (the edges of protocluster). The red star refers to the SMG HDF750.1. The yellow triangle refers to the QSO. The red and blue squares around galaxies refers to galaxies with stellar masses $\geq 2.0\times 10^{9}M_{\odot}$ and $< 2.0\times 10
 ^{9}M_{\odot}$, respectively. The solid blue lines correspond to the escape velocity in an NFW halo for the central bin. The solid horizontal line at $\Delta v/\sigma$=0, corresponds to the systemic velocity of the protocluster.}
  
  \label{phase}
\end{figure}

In an attempt to further understand the dynamical state of this overdensity we complete our analysis showing in Figure~\ref{phase} the position of the confirmed members in the phase-space diagram, \citep[e.g.,][]{Oman13,Jaffe15,Haines15}. This diagnostic was developed for virialised NFW halos (including assumptions about spatial symmetrical distribution) which generally is not the case for any galaxy protocluster. However, although all these assumptions for such sparse structures, composed of several independent infalling groups, might not be entirely correct, some authors (e.g \citet{Shimakawa14}) have made estimates of $R_{200}$ and $M_{200}$ for the core of protoclusters at $z\sim2.5$. Thus, at such overdense structures as in our case, it should be worthwhile to do this diagnostic which could provide interesting findings about our system.

The $x$ component of the diagram is the projected distance from the cluster centre (RA.,DEC.)=(12:36:55.92, +62:13:26.40), normalized in unit of $R_{200}$. The $y$ component is the velocity offset from the systematic velocity of the protocluster along the line of sight normalized by the velocity dispersion of the protocluster. In particular, we use
\begin{eqnarray}
   R_{proj}=\sqrt{(x_{i}-x_{cl})^{2}+(y_{i}-y_{cl})^{2}}
\end{eqnarray}
\begin{eqnarray}
   \frac{\Delta v_{i}}{\sigma_{cl}}=\frac{c(z_{i}-z_{cl})}{(1+z_{cl})\sigma_{cl}}
\end{eqnarray}
where ($x$, $y$, $z$) are the two spatial components and the component in redshift space and c is the speed of light. The blue solid lines correspond to the escape velocity of the cluster assuming a \citet{Navarro96}
halo density profile. We follow \citet{Rhee17} and we calculate 
\begin{eqnarray}
   v_{esc}=\sqrt{\frac{2GM_{200}}{R_{200}}K(s)}
\end{eqnarray}
where
\begin{eqnarray}
   K(s)=\frac{\ln(1+Cs)}{s}g(C)
\end{eqnarray}
\begin{eqnarray}
   s=\frac{r_{3D}}{R_{200}}
\end{eqnarray}
\begin{eqnarray}
   g(C)=\left [\ln(1+C)-\frac{C}{1+C}\right ]^{-1}
\end{eqnarray}
$C$ is the concentration parameter, fixed at $C=6$ which is a typical value for cluster mass NFW halos \citep{Gill04}. The area within the dashed lines represents the area under the influence of the protocluster potential. In determining the velocity dispersion to compute $R_{200}$ and $M_{200}$ in Equation (7) we assume that only the 18 galaxies in the central bin $5.18\leq z\leq 5.208$ satisfy the NFW conditions. This choice ensures a clean sample with tight redshift distribution ($\Delta z$=0.028) and enables us to not overestimate the velocity dispersion getting a broader distribution of objects in redshift space. Thus, following the method of \citet{Beers90} and excluding the galaxies at the edges of protocluster (in the lower and higher redshift bin), we find that the velocity dispersion of the 18 sources is $\sigma_{cl}=1260 km/s$. We also separate the galaxies, according to their stellar mass. We separate the galaxies in two groups. The `red' group with $M_{\star}\geq 2.0\times10^{9}M_{\odot}$ and the `blue' one with $M_{\star}< 2.0\times10^{9}M_{\odot}$. Considering the schematic view of the orbit of a galaxy in phase space diagram, we can distinguish several regions according to \citet{Jaffe15}. The SMG lies at a small projected protocluster centric radii ($R_{proj}/R_{200}\sim1$) and has a velocity that is remarkably similar to the sistemic velocity of the protocluster ($\Delta v/\sigma \sim0$), indicating that it might be settling into the BCG of the future $z=0$ cluster. A group of `red' galaxies is close to the central region ($R_{proj}/R_{200}<2$) and have small velocities. They are approaching the virialised region from the right side. At $R_{proj}/R_{200}>2$ there is a number of `blue' galaxies that are likely in groups in the outskirts of protocluster, infalling in the main structure. We also show in the phase-space diagram the position of galaxies at the edge of the redshift distribution which we excluded in the $v_{esc}$ calculation. They have velocities close to the $v_{esc}$ and within the region enclosed by these two curves. It is very likely that they will remain bound by the massive halo of dark matter.

\section{Overdensity mass calculations}
In our analysis we have paid particular attention to the complex multi-component system in formation at $z=5.2$. In order to put this protocluster into an evolutionary context and to compare it with other known high-z protoclusters and with the cosmological simulations, the goal of this section is to explore the halo mass of the whole structure and the descendant mass at $z=0$. 

\subsection{Halo mass}
We derive the halo mass at $z=5.2$ with three different methods following \citet{Long20}. First of all, we estimate the total halo mass of protocluster by summing the individual halo associated to each galaxy and by using the stellar-to-halo abundance matching relationship presented in \citet{Behroozi13}. This method implicitly assumes galaxies as self-bound entities close to virialisation. We excluded the galaxies \#8 and \#4 as we do not have any estimate of their stellar masses. We estimate the M$_{DMH}$ ranging from 4.17$\times$10$^{10}$M$_{\odot}$ to 3.49$\times$10$^{12}$M$_{\odot}$ leading to a total halo mass of M$_{1,tot}$=7.95$\times$10$^{12}$M$_{\odot}$. The second method consists in identifying the central galaxy and using its stellar mass. As we do not know the stellar mass of the \#8, which corresponds to the SMG, we consider the most massive galaxy of our sample. Interpolating  this value over the `Behroozi-relationship' we find M$_{2,tot}$=1.9$\times$10$^{12}$M$_{\odot}$. 
Finally, in the last approach we sum up all masses of the individual cluster members to the total stellar mass of this  system.  We assume a fraction of 5\% of the baryonic to dark matter halo mass \citep{Behroozi18}. Thus we estimate a halo mass of M$_{3,tot}$=2.0$\times$10$^{12}$M$_{\odot}$. Assuming a halo mass of M$_{tot}\approx2-8\times$10$^{12}$M$_{\odot}$ already in place at redshift $z=5.2$, and simulated evolutionary tracks of protostructures from \citet{Chiang13} shown in Figure~7 in \citet{Long20}, the descendent mass in the local Universe is expected to be similar to a Virgo/Coma-like cluster. However, we do not include in the calculation the SMG (\#8) and a LAE (\#4). As a consequence, the corresponding halo mass has to be considered as a lower limit.

\subsection{Present day mass}
The method we use for computing the present day mass $M_{z=0}$  of this overdensity assumes a spherical collapse model \citep{steidel1998} where everything within the volume will collapse into a cluster. We first define the density contrast of our rest-frame UV bright sources (from now on UVgal) such as our LAEs and LBGs as
\begin{eqnarray}
   \delta_{gal}=\frac{n_{UVgal}-\bar{n}_{UVgal}}{\bar{n}_{UVgal}}
\end{eqnarray}
where $n_{UVgal}$ is the number density of rest-frame UV bright sources within the region under consideration and $\bar{n}_{UVgal}$ is the mean number density of rest-frame UV bright sources in the field. We obtain the number densities of this source population by simply dividing the observed number counts of these sources in a given area. The spectroscopic sample for this analysis includes the sample of sources discovered by \citet{Walter12}. However, we will only consider the objects within the OSIRIS FoV. We find that 22 galaxies lie within the FoV area of 45 arcmin$^{2}$ (17$\times$14 cMpc$^{2}$). The number density is  $n_{UVgal}$=0.49$\pm$0.10 arcmin$^{-2}$. The uncertainty for the density measurement is calculated assuming the Poissonian noise  $\sigma_{N}=\frac{\sqrt N}{area}$ where N is the number of galaxies. We use as field number density the density published by \citet{Bouwens15}. At $z\sim5$ the $\bar{n}_{LAEs}$=0.2771$\pm$0.0194 arcmin$^{-2}$. With these values, we find the number density contrast $\delta_{gal}$=0.77$\pm$0.27. The 2D density contrasts between protoclusters and the field are relatively small at high redshift. Because the number density by \citet{Bouwens15} is calculated for a dataset of bright UV LBG galaxies, the estimated $\delta_{gal}$ has to be considered as an upper limit. To compute the mass of the overdensity contained in the volume occupied by the overdensity, we use the classical equation presented by \citet{steidel1998}
\begin{eqnarray}
   M_{z=0}=\langle \rho \rangle V_{true}(1+\delta_{m})
\end{eqnarray}
where $\langle \rho \rangle$ is the mean matter density of the Universe at $z=5.2$, $V_{true}=\frac{V_{obs}}{C}$ is the distortion corrected comoving observed volume and $\delta_{m}$ is the matter density contrast.
The value of $\delta_{m}$ is expressed by the equation
\begin{eqnarray}
   1+b\delta_{m}=C(1+\delta_{gal})
\end{eqnarray}
where $b$ is the bias parameter and C the correction factor for the redshift space distortion \citep{Kaiser87,steidel1998,Overzier16}. Applying a linear interpolation to the relation found by \citet{Ouchi18}, we compute the bias parameter at redshift $z=5.2$ to be $b$ = 3.45. The matter density contrast $\delta_{m}$ is interconnected to the C factor by the equation
\begin{eqnarray}
   C=1+f-f(1+\delta_{m})^{\frac{1}{3}}
\end{eqnarray}
where $f$ is a function of redshift and depends on the cosmological model \citep{Linder05,steidel1998}
\begin{eqnarray}
   f(z)=\Omega_{m}(z)^{0.6}
\end{eqnarray}
which we take to be $\simeq0.99$ at $z=5.2$. 
Solving simultaneously the Equation (14) and (15) in $\delta_{m}$ and C, we derive
\begin{eqnarray}
   C=0.94   \qquad  \delta_{m}=0.19
\end{eqnarray}
To estimate the observed volume, we multiplied the spatial extent and the sky projected area. In the line of sight dimension, the effective spatial extent is represented by the
difference in the radial comoving distance between the two
redshift boundaries covered (from z=5.155
to 5.224). The range of $\Delta z=0.069$ is a factor $\sim 2$ smaller than the redshift range probed by the medium-band filter ($\Delta z =0.132$). Setting the volume (in redshift space) containing the overdensity as $\sim 17\times14\times34.8$ cMpc$^{3}$, the corrected survey volume $V_{true}$=8,811 cMpc$^{3}$ and the mean matter density of the Universe at $z=5.2$ $\langle \rho \rangle$=4.1$\times$10$^{10}$M$_{\odot}$cMpc$^{-3}$, we compute a lower limit of the mass of the overdensity from Equation (13) that is $M_{z=0}$=4.3$\times$10$^{14}$M$_{\odot}$. The mass falls in the range typical of a Virgo type cluster, i.e. 3-10$\times$10$^{14}$M$_{\odot}$. We can compare our present day mass estimate with the prediction of simulations for a protocluster in literature associated with the radio galaxy TN J0924–2201 at $z=5.2$ \citep{Venemans04,Overzier06}. For a window of 7$\times$7 arcmin$^{2}$ and $\Delta z\sim 0.07$, similar to our windows of observation, and galaxies with $SFR>1M_{\odot} yr^{-1}$, \citet{Chiang13} determined $\delta_{gal}\sim 1.5^{+1.6}_{-1.0}$. Tuning the simulation predictions to this particular observational configuration, the present day mass for this protocluster would be $M_{2201,z=0}$=4-9$\times$10$^{14}$M$_{\odot}$, consistent with our result. 

Basing on the assumption about the $\delta_{gal}$, the value of $\delta_{m}$ should be considered an upper limit. The typical value for a protocluster at z=5 is $\sim 0.2-0.4$ \citep{Suwa06}. Considering this value of $\delta_{m}$=0.19, in a spherical collapse model \citep{Mo96}, this is related to a linear matter enhancement of $\delta_{L}(z=0)=\delta_{m}(z=5.2)\times 5 \sim 0.95$ (where 5 is the linear growth factor from z=5.2 to z=0). This does not exceeds the collapse threshold of $\delta_{c}=1.68$. Therefore, the protocluster is expected not to collapse and virialise as a whole by now (z=0), suggesting that only the main protocluster progenitor grows into a Virgo-like cluster at the present day, while the clumps are going to evolve separately into independent halos. However, we are possibly observing only a part of a more complex and much larger structure. Generally, the size of protoclusters are larger than the area covered by one OSIRIS MOS pointing (FoV). Sizes of such structures are estimates to be up to 30~arcmin to 1~degree at the sky \citep{Muldrew2015,Casey16}. In section 7, we will discuss our results also in the light of theoretical predictions.

\section{The fate of the protocluster}
\subsection{General remarks}
One of the main challenges when protoclusters are revealed is to interpret the observations and place them into an evolutionary context that can predict their growth. The cosmic build-up starts in filaments, the Universe's backbone. This complex web of interconnected patterns shows a large variety of structures and substructures, of different scales and densities, at their intersections \citep{Evrard02}. However, a serious prediction of the fate of such structures is not easy, considering that their evolutionary history is a continuous adaptation in response to all environmental changes they experience. As a consequence, the different parameters that characterise the cosmic history of a protocluster depends on many components (e.g., dark matter, intracluster gas, environment imprints). Theoretically, the member galaxies of high redshift protoclusters occupy individual dark matter halos at the epoch at which they are being observed but later they will merge into a common halo by $z=0$. The puzzle we try to examine in this section is to understand whether the future fate of galaxies within a newly confirmed overdensity at $z\sim5.2$ will be in a single collapsed halo or the individual density peaks will evolve in independent halos as part of a supercluster. It is important to point out that the results of the analysis we carried out, using our spectroscopic observations and literature data \citep{Walter12}, do not cover the full area of the overdensity as traced by the parent candidate member sample of 55 LAEs and LGBs found by \citet{Arrabal18}. We note that the SHARDS region in GOODS-N has a size of about 130 sq. arcmin \citep{Perez13}. However, although we cannot establish the spatial extent of the protocluster and a definitive evolutionary framework, we try to address this question by examining the key clues we found. 

\subsection{A collapsed system or a supercluster with filaments, bridges?}
As stated in the previous sections, we find a very extended protocluster, 34.8~cMpc ($\sim$5.6 Mpc physical size) with an elongated shape from NE to SW, formed by multiple components linked through filaments clearly shown in the redshift distribution. Sizes of high-redshift protoclusters (as progenitors of local galaxy clusters) have been widely studied by using especially numerical simulations, depending significantly on the resultant halo mass at $z=0$. \citet{Suwa06} at $z=4-5$ found that the progenitors are extended regions of typically 20-40~cMpc with dark matter halos in excess of $10^{12} h^{-1}M_{\odot}$. \citet{Chiang13} defined an effective radius $R_{e}$ which enclose 40\% of the total mass of a protocluster, estimating that the typical diameter $2R_{e}$ of a protocluster at $z\sim5$ is of $2R_{e}$=13.2$^{+2.8}_{-2.4}$~cMpc and $2R_{e}$=18.8$^{+3.2}_{-3.2}$~cMpc for progenitors of Virgo-type clusters ($(3-10)\times 10^{14}M_{\odot}$) and Coma-type clusters ($>10^{15}M_{\odot}$), respectively. Using a different measure of the protocluster radius \citet{Muldrew15} found that the average radius that encloses 90\% of the stellar mass of a protocluster at $z\sim5$ is $\sim 2-3(4)$~Mpc in physical scale for final cluster masses of $(1-9)\times 10^{14}M_{\odot}(>10^{15}M_{\odot})$. We identify in our protocluster four substructures around the central region which hosts the SMG. The NE region contains an off-centre core, whose galaxies are members of a pair-like system with a dynamical mass M$_{200}\sim$9.4$\times$ 10$^{12}$M$_{\odot}$. From the clustering analysis, we identify this region as the most evolved region. This might be evidence of a 
`proto-red sequence' in formation. The presence of a pair-like subgroup is statistically reliable in protoclusters, because they form faster in the core in agreement with the hierarchical structure formation model \citep{Toshikawa14,Toshikawa16}. \citet{Kubo15} predict that such rare groups hosted in massive halos with $M_{vir}= 10^{13.4}-10^{14.0}M_{\odot}$ might evolve and merge into the brightest cluster galaxies (BCGs) of the most massive clusters at present. The other three components identified in our structure, whose galaxies do not span the same physical volume as the galaxies in the NE region, are an indication of the young evolutionary state of this overdensity at $z=5.2$ around the the dusty starburst HDF850.1.

Similar large complex structures are rare in the distant Universe, but not unexpected \citep[see e.g.,][]{Muldrew15,Casey16}. For example, \citet{Dey16}  found two overdense regions at $z=3.786$, the protocluster PC~17.96+32.3 through Keck/DEIMOS observations. The small velocity dispersions of its subgroups and the spatial distribution suggest that these systems are dynamically young and in the process of merging. \citet{Topping16,Topping18} investigated the nature and evolution of large-scale structures within the SSA22 protocluster region at $z=3.09$ using both KECK/LIRS spectroscopic observations and simulations. They found that the observed double-peaked structure in the SSA22 redshift distribution corresponds not to a single coalescing cluster but rather to two different cluster progenitors with masses of $\sim {10}^{15}{h}^{-1}\,{M}_{\odot }$ and $> {10}^{14}{h}^{-1}\,{M}_{\odot }$. Recently, \citet{Toshikawa20} carried out optical follow-up spectroscopy on three overdense regions at $z=4.898$, 3.721 and 3.834 in the CFHTLS Deep Fields. They found that small groups in large assembly structures appear already at $z\sim4-5$ in parallel with the formation of the protocluster core. However, their evolution depends on the halo mass. Indeed, according to theoretical predictions, if the subgroups have quite a small redshift separation from the main protocluster component, and the protocluster is the progenitor of a significantly rich cluster ($>10^{15}M_{\odot}$) at $z=0$, it is likely possible that the neighbouring groups will be incorporated into a single halo by $z=0$. Alternatively, the subgroups will evolve in satellites of a supercluster by $z=0$. A massive multi-component supercluster called the 
`Hyperion' has been unveiled by \citet{Cucciati18} at $z\sim2.45$ in the COSMOS field using the spectroscopic VIMOS Ultra-Deep Survey (VUDS), complemented by the zCOSMOS-Deep spectroscopic sample. They found a complex structure which contains at least seven density peaks with masses in the range $\sim 0.1\times10^{14}M_{\odot}-2.7\times10^{14}M_{\odot}$ that are independently evolving and in process of collapsing. Several of these density peaks have been previously reported by e.g., \citet{Diener13,Casey15,Wang16}.

In the light of the above considerations and the detailed characterization of this overdensity, we can speculate about the fate of our massive protocluster at $z=5.2$. From the analysis of the velocity structure and spatial distribution of all components of this protocluster, we demonstrated the existence of a large scale structure around the SMG HDF850.1. The spatial extent and the complex structure shows that it is very extended in several directions. The neighboring clumps which belong to the edges of the structure and are connected to the highest overdense region through rope-like filaments, have small separations from the central core of this protocluster around the well-known SMG ($\Delta z< 0.04$). Considering our results and according to simulations \citep{Chiang13} and other known high redshift galaxy protoclusters \citep{Toshikawa20}, we can assess that the main progenitor of this structure is bound to collapse into a Virgo-like galaxy cluster at z=0 with neighboring independent evolving clumps, comparable with rather elongated filements typical of superclusters. However, our results are derived from a pilot spectroscopic program of a structure that could be incomplete in terms of areal coverage of the field. Additionally, all the mass estimates we performed are conservative estimates. Thus, we expect that a more complete census of the galaxies residing in the protocluster and its surroundings can reveal an evolution at $z=0$ similar to a more massive Coma-like cluster ($>10^{15}M_{\odot}$), which eventually incorporate the neighboring clumps into a single halo. Such a massive structure is not surprising at $z>5$ since we already have evidence of massive protoclusters at this epoch. \citet{Capak11} found at $z=5.3$ a cluster of massive galaxies which extends over $>13$ Mega-parsecs, and contains a luminous 
quasar as well as a system rich in molecular gas. At earlier epochs, \citet{Jiang18} found at $z=5.7$ a protocluster with at least 41 confirmed members. It occupies a volume of about 35 x 35 x 35 cubic co-moving Mpc and is predicted to collapse into a galaxy Coma-like cluster with a mass of $3.6\pm0.9\times10^{15}M_{\odot}$. In our pilot spectroscopical observations, we find strong evidence that this is one of the richest protoclusters at beyond redshift $z=5$.

\section{Conclusions}
We conducted a pilot spectroscopic follow-up with the multi-object spectrograph OSIRIS on the GTC of a sample of 17 rest-frame UV sources (LAEs and LGBs) selected from AH18. These sources are candidate members of the protocluster PCL$-$HDF850.1 at $z=5.2$. The major aim is to probe the existence of one of the largest known and most overdense high-redshift structure beyond redshift $z=5$. The main results of this work are summarised below:
 
    \spb We spectroscopically confirmed 13 cluster members, 10 LAEs and 3 LBGs, that are part of an already known overdensity found by \citet{Walter12} at redshift $z=5.2$. The objects span a redshift range between $5.155\leq z_{gal}\leq 5.224$. 3 of them match the 13 sources discovered by \citet{Walter12} while ten members are completely new. 
    
    \spb We investigate the properties of 13 members and we obtained spectra for them. The analysis of the Ly$\alpha$ equivalent widths and SFRs are consistent with starburst galaxies. Among the candidates we detect a powerful AGN, the quasar $\rm CXOHDFN J123647.9+620941$ \citep[already spectroscopically identified in the Chandra Deep Field-North by][]{Barger02,Barger03}). Except for the AGN, we cannot find any significant differences among the properties of these galaxies.

    \spb The dusty starburst HDF850.1 seems to evolve into the BCG of this cluster in formation. Interestingly, none of the confirmed members (beside the SMG) are bright in the far-infrared/submm/radio wavelength regime. The SFR of this structure seems to be dominated by the well-known dusty starburst.
    
    \spb We apply the FoF analysis on the 13 confirmed LAEs and LBGs at $z=5.2$ from this work, combined with the sample of 10 members from \citet{Walter12} in order to assess the possibility of bounded regions. All 23 sources span a comoving distance of 34.8~cMpc (diameter). From the histogram of the redshift distribution and the clustering analysis we find a clear segregation of galaxies along the structure of the protocluster. In the NE part we find a region with ten galaxies, nine in a narrow redshift range between $5.183\leq z\leq 5.208$ and one at $z=5.163$. We define this region as Clump1 with an `off-centre core'. It could also represent the 
   `proto-red sequence' of this overdensity. In the southern part of this structure, we find three additional clumps with $\Delta z<0.04$, which surround the central core with the SMG HDF850.1. They could evolve independently from the main protocluster.
    
    \spb Most probably, our spectroscopic observation (one OSIRIS pointing) do not cover the full area of the overdensity. We put this protocluster into an evolutionary framework and predict its fate comparing our results with other known high-z protoclusters and with simulations. We find that the size of the protocluster and the estimate of the mass using the spherical collapse model are consistent with simulations and suggest that this structure does not collapse as a whole but the main progenitor will evolve into a cluster of mass $\sim 4.3 \times 10^{14}M_{\odot}$ by $z=0$ similar to a Virgo-cluster, with independent satellite halos. The latter will possibly be incorporated in a single halo if a more detaliled analysis of this impressive structure indicates that the protocluster is the progenitor of a more massive Coma-like cluster ($>10^{15}M_{\odot}$).
\\[12pt]
Based on the results of this pilot program, we prove the existence of a rich overdensity at $z=5.2$. Considering that in AH18 we have photometrically identified a total of 44 potential new members of this overdensity, with this program we observed less than 50\% of this impressive large structure. Thus, future GTC OSIRIS observations will aim to fully characterise this rich protocluster with all potential members.  

\section*{Acknowledgments} We thank the referee for his/her useful comments, which allowed us to clarify some parts of the paper. H.D. acknowledges financial support from the Spanish Ministry of Science, Innovation and Universities (MICIU) under the 2014 Ramón y Cajal program RYC-2014-15686 and AYA2017-84061-P, the later one co-financed by FEDER (European Regional Development Funds). R.C., P.A.H., J.M.R.E. and C.M.T.  want  to  acknowledge  support  from  the  Spanish  Ministry  Science, Innovation and Universities (MICIU) under grants AYA2015-70498-C2-1-R, AYA2013-47742-C4-2-P and AYA2016-79724-C4-2-P. P.G.P acknowledges support from Spanish Government grant PGC2018-093499-B-I00. Based on observations made with the Gran Telescopio Canarias (GTC), installed in the Spanish Observatorio del Roque de los Muchachos of the Instituto de Astrof\'{i}sica de Canarias, in the island of La Palma. We would like to thank Emanuele Daddi, Jos\'{e} Eduardo M\'{e}ndez Delgado, Daizhong Liu, Arianna Long, Rodrik Obverzier, Jose\'{e} P\'{e}rez Mart\'{i}nez and Andrea Negri for valuable help to specific issues of this manuscript.

\section*{Data Availability}
The raw data underlying this article are available in the GTC public archive at https://gtc.sdc.cab.inta-csic.es/gtc/ and can be accessed with the proposal number GTC122-17B. The reduced data underlying this article will be shared on reasonable request to the corresponding author
\bibliographystyle{mnras}
\interlinepenalty=10000
\bibliography{biblio}
\label{lastpage}
\end{document}